\documentclass[pra,aps,showpacs,groupedaddress,superscriptaddress,twocolumn,toc=flat]{revtex4-1}

\usepackage[colorlinks=true,linkcolor=blue,urlcolor=blue,citecolor=blue]{hyperref}

\usepackage[utf8x]{inputenc}
\usepackage{color}
\usepackage{bbm} 

\usepackage{amsfonts,amsmath,amssymb,stmaryrd}

\usepackage{graphicx}
\usepackage{subfigure}  
\usepackage{bbm} 
\usepackage{hyperref}
\usepackage{epsfig}
\usepackage{mathrsfs}
\usepackage{verbatim}
\usepackage{centernot}
\usepackage{ulem}
\usepackage{nicefrac}

\renewcommand{\l}{\left(}
\renewcommand{\r}{\right)}

\newcommand{\bra}[1]{\langle#1|}
\newcommand{\ket}[1]{|#1\rangle}

\renewcommand{\ij}{{\langle \vec{i}, \vec{j} \rangle}}

\renewcommand{\H}{\hat{\mathcal{H}}}

\renewcommand{\c}{\hat{c}}
\renewcommand{\a}{\hat{a}}

\newcommand{\cd}{\hat{c}^\dagger}

\newcommand{\ad}{\hat{a}^\dagger}

\renewcommand{\b}{\hat{b}}
\newcommand{\hd}{\hat{h}^\dagger}
\newcommand{\h}{\hat{h}}

\newcommand{\hc}{\text{h.c.}}

\newcommand{\f}{\hat{f}}
\newcommand{\fd}{\hat{f}^\dagger}

\usepackage{array}

\usepackage{cancel,ifthen}
\newcommand{\cmnt}[2][NoInPuT]{\ifthenelse{\equal{#1}{NoInPuT}}{}{{\color{red}\sout{#1}}} {\color{blue} #2}}

\usepackage{bm}	
\renewcommand{\vec}[1]{\bm{#1}}

\begin{document}
\normalem	

\title{\textsf{\textbf{\Large Parton theory of ARPES spectra in anti-ferromagnetic Mott insulators}}}

\author{Annabelle Bohrdt}
\email[Corresponding author email: ]{annabelle.bohrdt@tum.de}
\affiliation{Department of Physics and Institute for Advanced Study, Technical University of Munich, 85748 Garching, Germany}
\affiliation{Munich Center for Quantum Science and Technology (MCQST), Schellingstr. 4, D-80799 M\"unchen, Germany}

\author{Eugene Demler}
\affiliation{Department of Physics, Harvard University, Cambridge, Massachusetts 02138, USA}

\author{Frank Pollmann}
\affiliation{Department of Physics and Institute for Advanced Study, Technical University of Munich, 85748 Garching, Germany}
\affiliation{Munich Center for Quantum Science and Technology (MCQST), Schellingstr. 4, D-80799 M\"unchen, Germany}

\author{Michael Knap}
\affiliation{Department of Physics and Institute for Advanced Study, Technical University of Munich, 85748 Garching, Germany}
\affiliation{Munich Center for Quantum Science and Technology (MCQST), Schellingstr. 4, D-80799 M\"unchen, Germany}

\author{Fabian Grusdt}
\affiliation{Department of Physics and Arnold Sommerfeld Center for Theoretical Physics (ASC), Ludwig-Maximilians-Universit\"at M\"unchen, Theresienstr. 37, M\"unchen D-80333, Germany}
\affiliation{Department of Physics and Institute for Advanced Study, Technical University of Munich, 85748 Garching, Germany}
\affiliation{Munich Center for Quantum Science and Technology (MCQST), Schellingstr. 4, D-80799 M\"unchen, Germany}

\pacs{}

\date{\today}

\begin{abstract}
Angle-resolved photoemission spectroscopy (ARPES) has revealed peculiar properties of mobile dopants in correlated anti-ferromagnets (AFMs). But describing them theoretically, even in simplified toy models, remains a challenge. Here we study ARPES spectra of a single mobile hole in the $t-J$ model. Recent progress in the microscopic description of mobile dopants allows us to use a geometric decoupling of spin and charge fluctuations at strong couplings, from which we conjecture a one-to-one relation of the one-dopant spectral function and the spectrum of a constituting spinon in the \emph{undoped} parent AFM. We thoroughly test this hypothesis for a single hole doped into a 2D Heisenberg AFM by comparing our semi-analytical predictions to previous quantum Monte Carlo results and our large-scale time-dependent matrix product state (td-MPS) calculations of the spectral function. Our conclusion is supported by a microscopic trial wavefuntion describing spinon-chargon bound states, which captures the momentum and $t/J$ dependence of the quasiparticle residue. Our conjecture suggests that ARPES measurements in the pseudogap phase of cuprates can directly reveal the Dirac-fermion nature of the constituting spinons. Specifically, we demonstrate that our trial wavefunction provides a microscopic explanation for the sudden drop of spectral weight around the nodal point associated with the formation of Fermi arcs, assuming that additional frustration suppresses long-range AFM ordering. We benchmark our results by studying the cross-over from two to one dimension, where spinons and chargons are confined and deconfined respectively.
\end{abstract}

\maketitle

~ \\
\textsf{\textbf{\large Introduction}}\\
The ARPES \cite{Damascelli2003} spectra of doped anti-ferromagnets (AFMs) have attracted considerable attention. In quasi 1D settings, they have revealed spin-charge separation: Instead of discrete delta-function peaks, a broad continuum signifies the existence of separate branches of a spin-less holon and a charge-neutral spinon \cite{Luttinger1963,Lieb1968,Ogata1990,Szczepanski1990,Weng1995,Kim1996,Kim1997,Bannister2000,Sing2003,Giamarchi2003}. The situation is strikingly different in the 2D Heisenberg AFM, the parent compound of high-$T_c$ cuprate superconductors \cite{Lee2006}. There, a discrete quasiparticle peak is found in the one-hole ARPES spectrum \cite{Wells1995,Ronning2005,Graf2007}, corresponding to a long-lived magnetic polaron \cite{SchmittRink1988,Kane1989,Sachdev1989,Elser1990,Dagotto1990,Martinez1991,Auerbach1991,Liu1992,Boninsegni1992a,Boninsegni1992,Leung1995,Brunner2000,Mishchenko2001,White2001,Manousakis2007,Mezzacapo2011,Koepsell2019,Grusdt2019PRB}. For reconciling the experimental observations with numerical calculations in the clean $t-J$ or Hubbard models, inclusion of electron-phonon interactions has been an important issue \cite{Cuk2005,Kar2008}. At finite doping, but before the system becomes superconducting, a pseudogap is observed \cite{Lee2008a}. Instead of a closed Fermi surface, as might be expected from a Fermi-liquid state, Fermi-arcs have been found at low energies around the nodal points $(\pm \pi/2 , \pm \pi/2)$ \cite{Shen2005} (we use units where the lattice constant $a=1$ and $\hbar = 1$). These arcs of high spectral weight appear like a part of a small Fermi surface, but the backside of the putative Fermi surface is invisible. The microscopic origin of Fermi arcs in the pseudogap phase of cuprates is not understood today, but their existence has been argued to imply exotic underlying physics and topological order \cite{Luttinger1960,Oshikawa2000,Chowdhury2015,Sachdev2016}.

Theoretically predicting ARPES spectra of real solids is challenging. Microscopic models are hard to solve because they involve non-trivial band structures, electron-phonon and electron-electron interactions; Moreover, model parameters are not exactly known. This has lead to a long-standing debate about the explanation of ARPES spectra in the undoped AFM insulator and the origin of Fermi arcs.

\begin{figure}[t!]
\centering
\epsfig{file=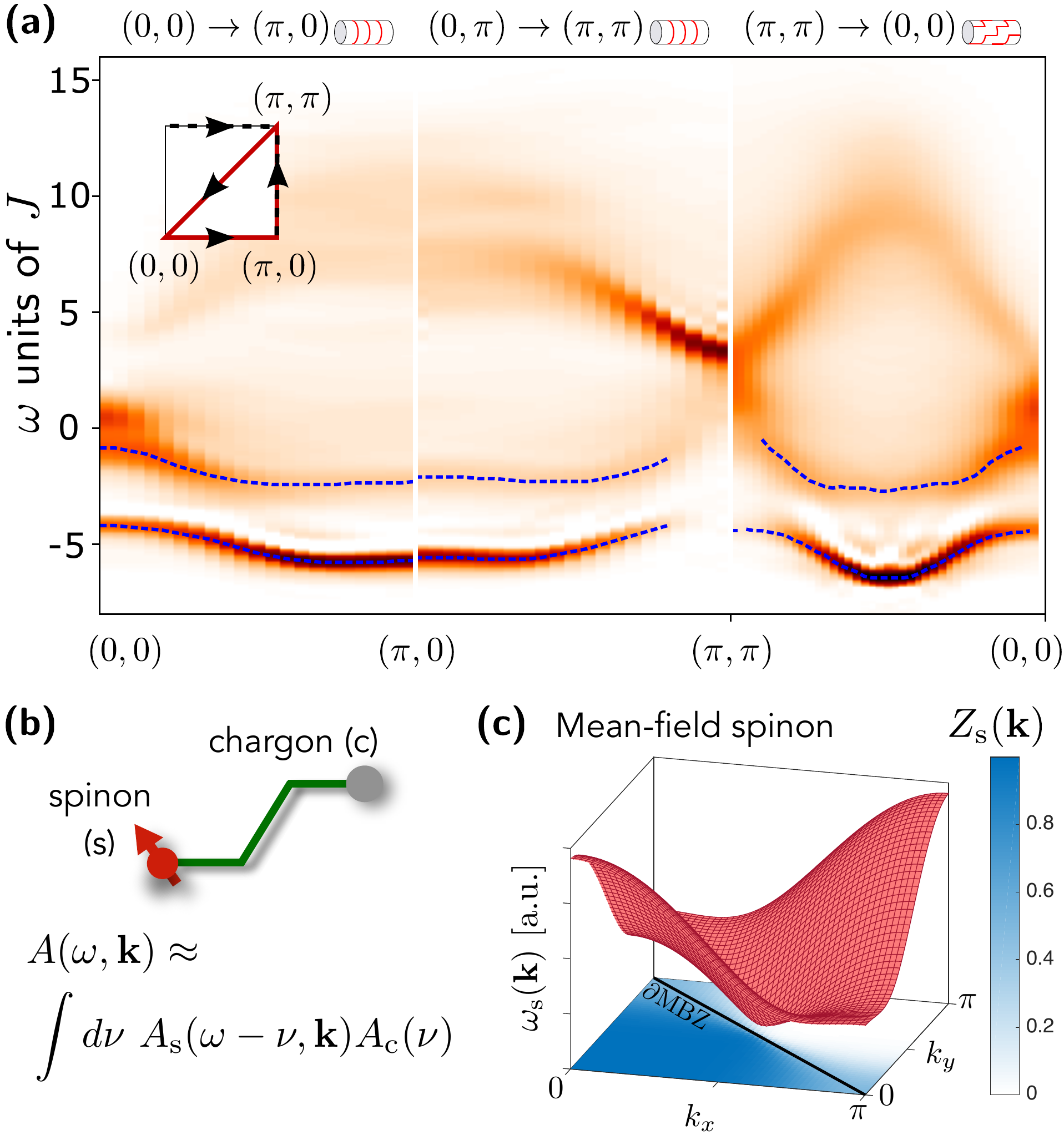, width=0.49\textwidth}
\caption{\textsf{\textbf{Magnetic polaron spectra and their unified description.}} (a) We perform td-MPS simulations of single-hole spectra in the $t-J$ model on $4 \times 40$ cylinders of different geometry. At strong couplings $t \gg J$, here $t=3 J$, a strong suppression of spectral weight is observed at $(\pi,\pi)$ at low-to-intermediate energies. Details of the td-MPS calculations are provided in part A) of the Methods. The spectrum is obtained along cuts in the Brillouin zone, calculated for different coverings of the cylinder by MPSs (both indicated in the top row). The dashed lines indicate the dispersion relations of the lowest two peaks (determined as local maxima of the spectrum), which we interpret as the ground and first vibrational states of the magnetic polaron. (b) At strong couplings, magnetic polarons can be understood as meson-like spinon-chargon pairs connected by geometric strings. The spectral function $A(\omega,\vec{k})$ of a hole can be approximated by a convolution of the spinon and chargon contributions $A_{\rm s}(\omega,\vec{k})$ and $A_{\rm c}(\omega)$ respectively, where the center-of-mass momentum of the meson is carried by the heavy spinon. (c) The optimized slave-particle mean-field theory of fermionic spinons \cite{Lee1988,Piazza2015} yields the correct shape of the magnetic polaron dispersion $\omega_{\rm s}(\vec{k})$, with a minimum at the nodal point $(\pi/2,\pi/2)$ and low-energy states along the edge $\partial{\rm MBZ}$ of the magnetic Brillouin zone. The contribution to the spectral weight $Z_{\rm s}^{\rm MF}(\vec{k})$ predicted by the mean-field spinon ansatz is indicated by the color plot at the bottom. It features a sharp drop at the nodal point $(\pi/2,\pi/2)$, which may be the root of the missing spectral weight on the backside of the Fermi-arcs observed in cuprates \cite{Shen2005}.}
\label{figSummary}
\end{figure}

Here we focus on ARPES spectra in clean toy models for doped AFMs. Even in such idealized scenarios, the theoretical challenges are significant enough that many open questions remain and a universally accepted understanding is lacking. Our work contributes two significant advances: (i) we improve state-of-the-art numerical simulations of ARPES spectra and (ii) we combine our results with recent insights into the microscopic structure of charge carriers in doped AFMs \cite{Bohrdt2019NatPhys,Chiu2019Science} obtained from cold atom experiments with quantum gas microscopes. As a result, we reach a detailed understanding of one-hole ARPES spectra in the paradigmatic $t-J$ model.

Our findings have important consequences, both theoretically and experimentally. Ultracold atom experiments enable clean studies of the Fermi-Hubbard model with tunable parameters \cite{Hart2015,Greif2013,Cheuk2016,Boll2016,Mazurenko2017,Chiu2019Science,Koepsell2019}, in 1D, 2D, or in continuous dimensional cross-overs which are hard to realize in solids. ARPES spectra can be accessed in optical lattices \cite{Kollath2007,Stewart2008,Greif2011,Torma2016,Bohrdt2018,Brown2019}, allowing to experimentally test our theoretical predictions in the near future. On the theoretical side, our work verifies that one-hole ARPES spectra in the AFM can be understood from more fundamental constituents (partons), whose properties we can describe on a quantitative and microscopic level. Moreover, this leads to new insights to the current puzzles of cuprates, in particular the microscopic origin of Fermi arcs. 

We perform microscopic numerical calculations and study spectral properties of magnetic polarons in the 2D $t-J$ model. On the one hand, we use unbiased time-dependent matrix product state (td-MPS) simulations \cite{Zaletel2015,Paeckel2019,Kjall2013} to calculate the one-hole ARPES spectrum on four-leg cylinders, see Fig.~\ref{figSummary} (a). Our work builds upon recent progress in the calculation of dynamical response functions using td-MPSs \cite{Gohlke2017,Verresen2018spec}. On the other hand, we use an analytic trial wavefunction \cite{Grusdt2019PRB} and show that it provides a complete physical picture of the observed low-energy features in the spectrum; see Fig.~\ref{figSummary} (b) and (c) for a summary.

The main results of our paper are as follows. First, we argue that state-of-the-art numerical calculations of the spectral function conclusively demonstrate that magnetic polarons in the clean $t-J$ model are composed of partons: they form meson-like bound states of spinons and chargons. Second, and in contrast to general wisdom, the spin-charge correlations present in this model at strong coupling can be efficiently described by a Born-Oppenheimer product wavefunction, \emph{if} one works in the so-called geometric string basis \cite{Grusdt2018PRX,Grusdt2019PRB}. As an important consequence of this second result, we demonstrate that all characteristic low-energy features in the spectrum at strong coupling can be attributed to either spinon or chargon properties. Third, we have a simple analytic understanding of the microscopic chargon properties. This leads us to the main conjecture of our work: namely, that \emph{a one-to-one relation exists, valid at strong coupling, between the observed one-hole spectral function and the spectrum of a constituting spinon in the undoped parent AFM}. This result has consequences well beyond the present work, suggesting ARPES spectroscopy at strong coupling as the most direct tool yet to probe the properties of constituting spinons in quantum AFMs. Possible applications include studies of quantum spin liquids.

Our paper is organized as follows. In the remainder of the introduction, we provide an overview of the main spectral features considered later. We introduce the model Hamiltonian and explain how our results relate to earlier studies. In the following section, we derive the parton theory of ARPES spectra in the geometric string basis. Then we present our td-MPS results, which contain new evidence for the parton nature of magnetic polarons. After establishing the known chargon features in the spectrum, we discuss the rich momentum dependence of the quasiparticle residue $Z(\vec{k})$ observed at strong couplings. We explain our observations by relating them to spinon properties -- which, in turn, we predict by an analytical trial wavefunction. We close by discussing how our findings may be related to Fermi arcs observed in cuprates, and how related types of experiments on quantum spin liquids can be analyzed in a similar way. 

\textbf{Model.} 
In the following we will consider the 2D $t-J$ model, defined by the Hamiltonian $\H=  \sum_{\mu=x,y} \H_t^\mu + \H_J^\mu$. It is believed to capture the essential low-energy physics of the anisotropic 2D Fermi-Hubbard model with on-site interaction $U$ and tunnelings $t_\mu$ in $\mu=x$ and $y$ directions \cite{Auerbach1998}. The individual terms are of tunneling type,
\begin{equation}
 \H_t^\mu = - t_\mu \sum_{\vec{j}}  \sum_\sigma \hat{\mathcal{P}}_{\rm GW}  \bigl( \cd_{\vec{j} + \vec{e}_\mu,\sigma} \c_{\vec{j},\sigma} + \hc \bigr) \hat{\mathcal{P}}_{\rm GW},
\label{eqHt}
\end{equation}
and AFM spin-exchange terms of strength $J_\mu = 4 t_\mu^2 / U$,
\begin{equation}
 \H_J^\mu =   J_\mu \sum_{\vec{j}}  \l \hat{\vec{S}}_{\vec{j}+\vec{e}_\mu} \cdot \hat{\vec{S}}_{\vec{j}} - \frac{\hat{n}_{\vec{j}+\vec{e}_\mu} \hat{n}_{\vec{j}} }{4} \r.
 \label{eqHJ}
\end{equation}
Here $\cd_{\vec{j},\sigma}$ creates a fermion with spin $\sigma$ on site $\vec{j}$, $\vec{e}_\mu$ denotes a unit vector along $\mu=x,y$, and $\hat{\mathcal{P}}_{\rm GW}$ is the Gutzwiller projector on a subspace with zero or one fermion per lattice site, $\hat{n}_{\vec{j}} = \sum_\sigma \cd_{\vec{j},\sigma} \c_{\vec{j},\sigma}=0,1$. We mostly focus on the case with exactly one hole, $\sum_{\vec{j}}\hat{n}_{\vec{j}} = L^2 - 1$, with $L$ the linear system size. If $\mu$ is not specified, $t$ and $J$ refer to the isotropic case $t=t_x=t_y$ and $J=J_x=J_y$. 

Throughout this paper we work in the strong coupling regime, where $t > J$, but before the Nagaoka polaron regime \cite{Nagaoka1966} is reached: $(t/J) < (t/J)_{\rm Nagaoka}$. Using large-scale DMRG \cite{White1993} simulations, White and Affleck \cite{White2001} have determined the critical value above which the Nagaoka polaron with a ferromagnetic core is realized, as $(t/J)_{\rm Nagaoka} = 40 \pm 10$.

\textbf{Overview and relation to previous works.} The magnetic polaron problem of a single hole moving in an AFM background is often considered to be essentially solved. Various semi-analytical and numerical techniques have been applied, and many of the key properties of magnetic polarons have been numerically established \cite{SchmittRink1988,Kane1989,Sachdev1989,Elser1990,Dagotto1990,Martinez1991,Auerbach1991,Liu1992,Boninsegni1992a,Boninsegni1992,Leung1995,Brunner2000,Mishchenko2001,White2001,Manousakis2007,Mezzacapo2011}. Nevertheless, there is no agreement on the correct physical interpretation of the obtained results. Partly, this can be attributed to conflicting numerical findings, and disconnected theoretical interpretations of the different features, as we explain next.

In the following, we summarize the main spectral features of a single hole in the 2D $t-J$ model, assuming $t>J$. We focus on low energies, no more than $\approx 2 t$ above the one-hole ground state. 
\begin{itemize}
\item[(i)] At the lowest energies, a dispersive quasiparticle peak -- the magnetic polaron -- is observed. Its bandwidth is on the order of the super-exchange coupling $J$ -- rather than hole tunneling $t$ -- and the shape of the dispersion relation differs significantly from that of a free hole. 
\item[(ii)] The quasiparticle residue $Z_{(\pi/2,\pi/2)}$ around the dispersion minimum at the nodal point depends strongly on $t/J$. All numerical methods have conclusively shown that $Z_{(\pi/2,\pi/2)} > 0$, despite conflicting theoretical proposals \cite{Sheng1996}.
\item[(iii)] Above the magnetic polaron ground state, at excitation energies $\Delta E < t$, a second peak has been observed. The most reliable signatures were obtained by Monte-Carlo calculations \cite{Brunner2000,Mishchenko2001}, while large-scale exact diagonalization studies yielded conflicting results for increasing system sizes \cite{Dagotto1990,Leung1995}. Like the ground state energy $E_0$ itself, the energy of the first peak $E_0$ has been shown to be consistent with a scaling of the form $E_n = - 2 \sqrt{3} t + c_n t^{1/3} J^{2/3}$, asymptotically for $t \gg J$.
\item[(iv)] The quasiparticle residue $Z(\vec{k})$ has strong, and non-monotonic momentum dependence. 
\end{itemize}
Our td-MPS studies confirm (i) and (ii); The resolution afforded by our method allows us to improve the predictions for the position of the first excited peak in (iii) and to study the dependence of $Z(\vec{k})$ in (iv) for larger values of $t/J$; Moreover, we numerically establish the following additional features, see Fig.~\ref{figSummary} (a):
\begin{itemize}
\item[(v)] Around $\vec{k}=(\pi,\pi)$ the spectral weight is suppressed in a wide window up to energies of order $\mathcal{O}(2t)$ above the ground state. 
\item[(vi)] The first excited peak -- see (iii) -- can be observed for all momenta, provided the ground state residue $Z(\vec{k})$ is non-negligible. The dispersion relation of the first excited peak is qualitatively identical to the ground state -- i.e. the excitation gap $\Delta_{\vec{k}}$ has only weak $\vec{k}$-dependence.
\end{itemize}

Previously, the following theoretical scenarios have been discussed:
\begin{itemize}
\item[(a)] \emph{String picture}: Early on, it has been proposed that strings of over-turned spins are attached to mobile dopants in a N\'eel state \cite{Bulaevskii1968,Brinkman1970,Trugman1988,Shraiman1988a,Manousakis2007,Golez2014}. This explains (iii), the scaling of the ground state energy $E_0 \simeq - 2 \sqrt{3} t + c_0 t^{1/3} J^{2/3}$ of a single hole at $t \gg J$; The string picture also predicts the existence of vibrationally excited states, whose energies should scale as $E_n = - 2 \sqrt{3} t + c_n t^{1/3} J^{2/3}$ -- in accordance with numerical observations \cite{Brunner2000,Mishchenko2001}. Recent ultracold atom experiments measured spin-spin \cite{Chiu2019Science,Bohrdt2019NatPhys} and spin-charge \cite{Koepsell2019} (see also \cite{Grusdt2019PRB}) correlation functions, which also support the string picture. Feature (ii) is also expected from the string picture, owing to the finite length of the strings. Features (i) and (iv) - (vi) require explanations beyond the string picture.
\item[(b)] \emph{Parton picture}: Based on phenomenological grounds and numerical evidence, B\'eran et al. \cite{Beran1996} proposed the parton picture, in which mobile dopants are described by fractionalized spinons and chargons. In a subsequent work \cite{Laughlin1997}, Laughlin drew an analogy with the 1D Fermi-Hubbard model and suggested that the low-energy ARPES spectrum in cuprates can also be interpreted in terms of point-like spinons and chargons, possibly interacting through a weakly attractive force. The parton picture explains (i): the dispersion relation of the one-hole ground state is determined by the spinon dispersion, which must have a bandwidth $W_{\rm s} = \mathcal{O}(J)$ dominated by spin-exchange. The conjectured chargon dispersion, with bandwidth $W_{\rm c} = \mathcal{O}(t)$, is expected to lead to additional features at higher energies in the spectrum. Features (ii) and (iii) are only consistent with the parton picture, if spinons and chargons form a bound state (they could be confined, or form a molecular bound state in a deconfined fractionalized Fermi liquid \cite{Senthil2003,Punk2015PNASS}). Scenarios with spin-charge separation as envisioned by Anderson \cite{Anderson1987}, with $Z = 0$ and as found in 1D, can be ruled out numerically \cite{Mishchenko2001} at infinitesimal doping. To make quantitative predictions and fully explain features (ii) - (vi), detailed knowledge about the parton dispersions and their microscopic interactions is required; this is typically beyond the scope of phenomenological descriptions. An experimental work \cite{Graf2007} has also led to an interpretation of the pronounced high-energy features in the spectrum as signatures of spinon and chargon branches.  
\item[(c)] \emph{Polaron picture}: The most widely used microscopic picture so far, has been the polaron scenario \cite{SchmittRink1988,Kane1989,Sachdev1989,Martinez1991,Auerbach1991,Liu1992,Boninsegni1992a,Boninsegni1992}. As the mobile dopant moves through the AFM, one assumes that it  interacts with collective magnon excitations. This picture should not be considered to be separate from (a) and (b): For example, strong interactions with magnons can describe strings of over-turned spins attached to the dopant. Spin-wave calculations of the spectral function \cite{Martinez1991,Liu1992} have revealed several vibrational peaks with the expected scaling $\simeq t^{1/3} J^{2/3}$ of their energies \cite{Liu1992}, thus explaining (iii). The strong renormalization of the bandwidth of the dopant (i), from $\mathcal{O}(t)$ to the observed $\mathcal{O}(J)$ is also predicted, although without identifying a clear physical mechanism. This is a general disadvantage of the polaron picture: when $t > J$ the system is so strongly coupled that all predictions require advanced numerics or uncontrolled approximations. While the polaron picture per se is certainly correct, it is of little help in the identification of simpler constituents of these polarons.
\end{itemize}

The goal of this article is to establish a unifying physical picture, which is able to explain the rich phenomenology (i) - (vi) of the strong coupling ARPES spectrum. We combine the parton and string pictures of magnetic polarons, by arguing that the latter are composed of spinons and chargons connected by universal (geometric) strings, see Fig.~\ref{figSummary} (b). Importantly, we provide quantitative descriptions of both ingredients, including a microscopic trial wavefunction \cite{Grusdt2018SciPost,Grusdt2019PRB}. Moreover, we explain why -- at strong coupling -- any feature of the spectrum is determined by \emph{either} the spinon \emph{or} the chargon / string properties: Essentially, a Born-Oppenheimer product ansatz in the geometric string basis \cite{Grusdt2018PRX,Grusdt2019PRB} allows us to factorize spinon and chargon contributions. 

As summarized above, the string picture explains features (ii) and (iii). Since we assume that $t \gg J$, we can first neglect the effects of spinon dynamics and describe strings independently. We will demonstrate that the observed strong $t/J$ dependence of the quasiparticle weight (ii) and the excitation energies (iii) can be explained, even quantitatively, by a simple and universal semi-analytical calculation. This detailed understanding of the chargon, or equivalently string, properties sets the stage for closer analysis of the spinon properties. 

As mentioned above, feature (i) naturally emerges in a parton description of magnetic polarons. Feature (vi) is a direct consequence of the product state nature of the spinon-chargon wavefunction: The first excited state is a string excitation but shares the same spinon properties as the ground state, including its dispersion relation. 

Here we go beyond earlier phenomenological studies of partons and demonstrate that \emph{quantitative} predictions of the spinon properties are possible. Our starting point is a parton theory of the \emph{undoped} Heisenberg AFM. Specifically, we focus on fermionic $U(1)$ Dirac spinons: These have previously lead to accurate variational predictions \cite{Lee1988,Piazza2015} (building upon Anderson's resonating valence bond paradigm \cite{Anderson1987,Baskaran1987}), and they have recently been proposed to provide a universal description of a larger class of quantum AFMs \cite{Song2019}. For example, the shape of the magnetic polaron dispersion, with its minimum at the nodal point, is inherited from the optimized spinon mean-field state of the Heisenberg AFM \cite{Grusdt2019PRB}. Similar observations were made in Refs.~\cite{Giamarchi1993,Wen1996}, but without including geometric strings which are necessary to describe, e.g., features (ii) and (iii).

The mean-field theory we use to describe spinons naturally predicts a strongly momentum dependent contribution $Z_{\rm s}(\vec{k})$ to the quasiparticle weight, see Fig.~\ref{figSummary} (c). Already on the mean-field level, a strong suppression of spectral weight around $(\pi,\pi)$ is predicted. Since the low-energy excited states of the magnetic polaron correspond to string excitations, sharing the same spinon contribution $Z_{\rm s}(\vec{k})$ to the quasiparticle weight as the ground state, the suppression of spectral weight around $(\pi,\pi)$ over a wide energy window is thus explained [feature (v)]. In this work we go beyond the mean-field theory, by including a Gutzwiller projection in our trial wavefunction. As a result, we find non-monotonic $\vec{k}$-dependence of $Z(\vec{k})$ for $t \gtrsim J$ -- explaining feature (iv), and in excellent agreement with unbiased numerical results. 

On the mean-field level, the spinon contribution to the quasiparticle weight $Z_{\rm s}(\vec{k})$ exhibits a sudden drop diagonally across the nodal point. This is a direct manifestation of the spinon Dirac cone and reminiscent of the phenomenology of Fermi arcs. In this paper we show that if the system has long-range N\'eel order and the $SU(2)$ symmetry is spontaneously broken, as in the optimized trial wavefunction \cite{Lee1988,Piazza2015} we use, the Gutzwiller projection in our magnetic polaron wavefunction widens the drop of $Z_{\rm s}(\vec{k})$ around the nodal point. However, we also show that a sharp drop survives the Gutzwiller projection, if $SU(2)$ invariance is restored in the trial state. This result goes beyond the scope of  mean-field parton theories. It may become relevant at finite doping in the $t-J$ model, when frustration restores the $SU(2)$ symmetry. In this regime we thus establish a possible \emph{microscopic} mechanism for the appearance of Fermi arcs, with strongly suppressed spectral weight on the backside of the Fermi pocket. Our microscopic results favor theoretical scenarios in which fermionic spinons and bosonic chargons are the effective constituents of the doped $t-J$ model.

~ \\
\textsf{\textbf{\large Results}}\\
\textbf{Parton theory of ARPES spectra.} 
We start by describing the general features of the ARPES spectrum expected from a parton theory of dopants in the 2D $t-J$ model. To simplify the single-occupancy condition built into Eq.~\eqref{eqHt} we introduce the parton representation,
\begin{equation}
\c_{\vec{j},\sigma} = \hd_{\vec{j}} \f_{\vec{j},\sigma}.
\label{eqDefSpnonHolon}
\end{equation}
Here $\h_{\vec{j}}$ is a chargon operator and $\f_{\vec{j},\sigma}$ denotes a $S=1/2$ spinon operator. The physical Hilbert space is defined by all states satisfying $\sum_{\sigma} \fd_{\vec{j},\sigma} \f_{\vec{j},\sigma} + \hd_{\vec{j}} \h_{\vec{j}} = 1$ for all $\vec{j}$. Notably, we do not yet have to specify the statistics of $\f$ and $\h$, respectively, at this point: both combinations (fermionic spinons and bosonic chargons / bosonic spinons and fermionic chargons) are allowed.

\emph{Spinon-chargon bound states at strong coupling.--}
We focus on the strong coupling limit $t \gg J$ of the isotropic model, where the fast motion of the hole can be approximately factorized in the geometric string basis \cite{Grusdt2018PRX,Chiu2019Science,Grusdt2019PRB}. We start from the state $\ket{\vec{j}^{\rm s}} \ket{0} = \sum_{\sigma} \c_{\vec{j}^s,\sigma} \ket{\Psi_0}$ where a hole is created by removing a fermion on site $\vec{j}^{\rm s}$ from the ground state $\ket{\Psi_0}$ of the undoped system. This state can be interpreted as a tightly bound state of the spinon and the chargon, both occupying the same site $\vec{j}^{\rm s}$. Next we include fast chargon fluctuations. If we properly account for the modified locations of the surrounding spins along the chargon trajectory, the back action on the spins by the chargon can be neglected if $t \gg J$. This so-called frozen-spin approximation (FSA) \cite{Grusdt2018SciPost,Chiu2019Science,Koepsell2019} has been shown to be very accurate if the undoped spin system has strong \emph{local} AFM correlations.

Concretely, we introduce an over-complete set of basis states \cite{Grusdt2019PRB}, $\ket{\vec{j}^{\rm s}} \ket{\Sigma}  = \hat{G}_\Sigma(\vec{j}^{\rm s}) \ket{\vec{j}^{\rm s}} \ket{0}$, where the operator $\hat{G}_\Sigma(\vec{j}^{\rm s})$ starts at site $\vec{j}^{\rm s}$ and translates the chargon and spins along the geometric string $\Sigma(\vec{j}^{\rm s})$:
\begin{equation}
\hat{G}_\Sigma(\vec{j}^{\rm s}) = \prod_{\ij \in \Sigma(\vec{j}^{\rm s})} \bigg( \hd_{\vec{i}} \h_{\vec{j}} \sum_{\tau=\uparrow,\downarrow} \fd_{\vec{j},\tau} \f_{\vec{i},\tau} \bigg).
\label{eqDefG}
\end{equation}
At strong coupling, $t \gg J$, spinon-chargon bound states with center-of-mass momentum $\vec{k}$ can approximately be described by
\begin{equation}
\ket{\Psi_{\rm sc}^{\rm FSA}(\vec{k})} = \frac{1}{L} \sum_{\vec{j}^{\rm s}} e^{i \vec{k} \cdot \vec{j}^{\rm s}} \sum_\Sigma \psi_\Sigma^{\rm FSA} \ket{\vec{j}^{\rm s}} \ket{\Sigma}.
\label{eqFSA}
\end{equation}
The FSA string wavefunction $\psi_\Sigma^{\rm FSA}$ ascribes complex amplitudes $\psi_\Sigma^{\rm FSA} \in \mathbb{C}$ to all string configurations $\Sigma$. The latter are independent of momentum $\vec{k}$ and can be calculated from an effective hopping model on the Bethe lattice, with an approximately linear string potential emanating from the spinon position $\vec{j}^{\rm s}$; see Refs.~\cite{Grusdt2018PRX,Grusdt2019PRB} for details. The state in Eq.~\eqref{eqFSA} describes a heavy spinon carrying momentum $\vec{k}$. The latter binds to itself the light chargon, which is delocalized over a large number of string configurations when $t \gg J$.

Now we will draw some general conclusions about the ARPES spectrum, assuming that it consists of spinon-chargon eigenstates described by the strong-coupling meson wavefunction in Eq.~\eqref{eqFSA}. As a further approximation, valid when the parent state $\ket{\Psi_0}$ has strong local AFM correlations, we assume that the basis states $\ket{\vec{j}^{\rm s}} \ket{\Sigma}$ are mutually orthonormal: $\bra{\vec{j}^{\rm s}\!~'} \vec{j}^{\rm s} \rangle \bra{\Sigma'} \Sigma \rangle \approx \delta_{\vec{j}^{\rm s}\! ~', \vec{j}^{\rm s}} \delta_{\Sigma',\Sigma}$. Otherwise, the following results do not depend on any specific parameters in Eq.~\eqref{eqFSA}. 

To calculate the spectrum, $A(\omega,\vec{k}) = {\rm Re} \frac{1}{\pi} \int_0^{\infty} dt ~ e^{i \omega t}$ $\times \bra{\Psi_0} e^{i \H t} \l \sum_\sigma \cd_{\vec{k},\sigma} \r e^{- i \H t} \l \sum_\sigma \c_{\vec{k},\sigma} \r \ket{\Psi_0}$, we note that the initial state on the right hand side is $\l \sum_\sigma \c_{\vec{k},\sigma} \r \ket{\Psi_0} = \ket{\vec{k}^{\rm s}} \ket{\Sigma = 0}$, where $\ket{\vec{k}^{\rm s}} = L^{-1} \sum_{\vec{j}^{\rm s}} e^{i \vec{k}^{\rm s} \cdot \vec{j}^{\rm s}} \ket{\vec{j}^{\rm s}}$ is a plane-wave spinon state. On the left hand side, $\bra{\Psi_0} e^{i \H t} = e^{i \omega_0 t} \bra{\Psi_0}$ reduces to a phase factor.

Because of the assumption that spinon-chargon states \eqref{eqFSA} are eigenstates, we can approximate $e^{- i \H t} \ket{\vec{k}^{\rm s}} \ket{0} \approx e^{- i \H_{\rm s} t} \ket{\vec{k}^{\rm s}} e^{- i \H_\Sigma t} \ket{0}$, where $\H_{\rm s}$ and $\H_\Sigma$ denote effective Hamiltonians of the spinon and string (chargon) respectively. We expect $\H_{\rm s} \propto J$ and $\H_{\Sigma} \propto t$, since these terms are dominated by spin-exchange and tunnel couplings respectively. Explicit forms of $\H_{\rm s}$ and $\H_{\Sigma}$ have been derived \cite{Grusdt2018PRX,Koepsell2019,Grusdt2019PRB}, and is has been demonstrated that they capture far-from equilibrium dynamics of a single hole on a remarkable quantitative level \cite{Grusdt2018PRX,Bohrdt2019Dyn} (see also Ref.~\cite{Hubig2019}).

As a result of the factorization of the eigenstates into spinon-chargon bound states, the spectral function becomes a convolution, 
\begin{equation}
A(\omega,\vec{k}) |_{\rm bound} = \int d\nu ~ A_{\rm s}(\omega - \nu,\vec{k}) A_{\rm c}(\nu).
\label{eqConv}
\end{equation}
The spinon contribution $A_{\rm s}(\omega ,\vec{k}^{\rm s}) = {\rm Re} \frac{1}{\pi} \int_0^{\infty} dt ~ e^{i \omega t}$ $\times \bra{\vec{k}^{\rm s}} e^{- i \H_{\rm s} t} \ket{\vec{k}^{\rm s}}$ depends on the momentum $\vec{k}^{\rm s}$ of the spinon. In contrast, the chargon contribution $A_c(\nu) = {\rm Re} \frac{1}{\pi} \int_0^{\infty} dt~ e^{i \nu  t} \bra{\Sigma=0} e^{- i \H_\Sigma t} \ket{\Sigma=0}$ is defined in the effective Hilbert space of geometric string states and has no $\vec{k}$ dependence. 

\begin{figure}[t!]
\centering
\epsfig{file=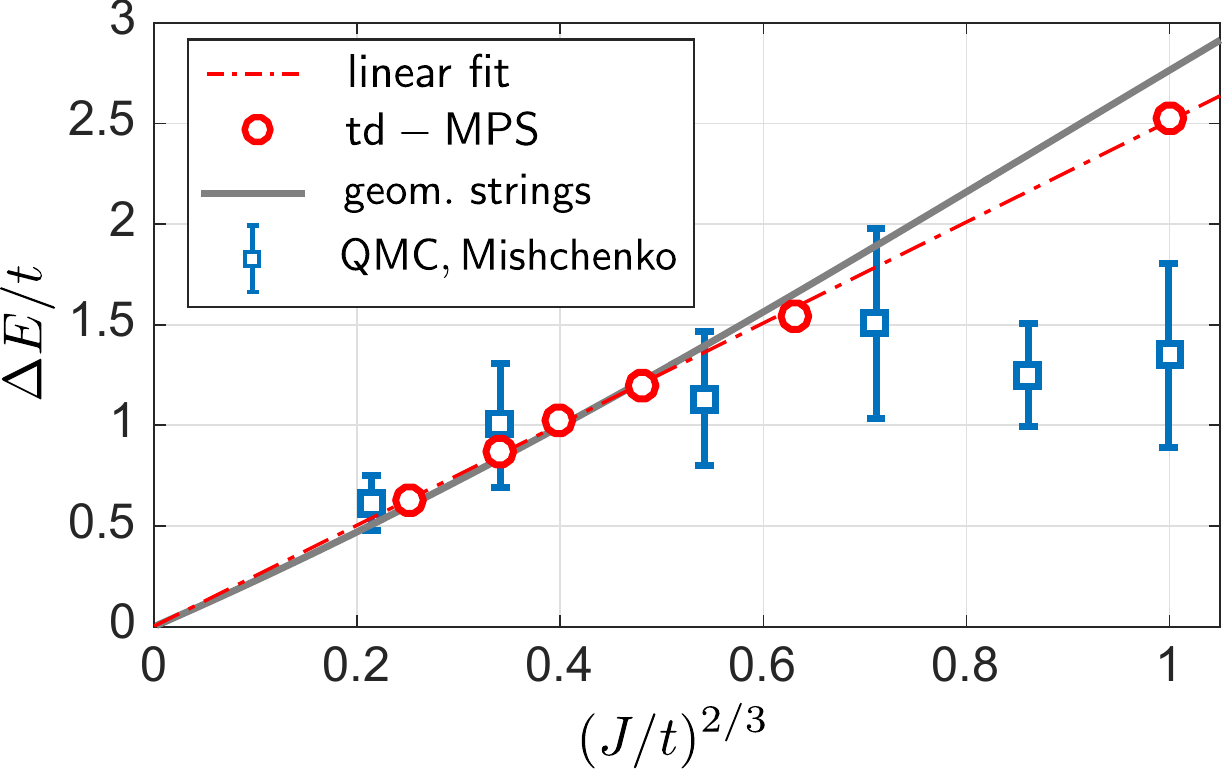, width=0.44\textwidth} $\quad$
\caption{\textsf{\textbf{Scaling of the first vibrational excitation.}} At strong coupling, $t>J$, we analyze the energy $\Delta E$ between the lowest two pronounced peaks in the spectrum. Our td-MPS results (red) on $4 \times L_x$ cylinders are compared to quantum Monte Carlo calculations by Mishchenko et al. (blue, data extracted from Ref.~\cite{Mishchenko2001}) and the effective geometric string approach (gray). A linear fit, $\Delta E / t = a (J/t)^{2/3} + b$, to our td-MPS data yields $a=2.51$ and $b=2 \times 10^{-3}$. All data is for the ground state at the nodal point, $\vec{k}=(\pi/2,\pi/2)$. See part A) of the Methods for a discussion how the peaks are extracted from our numerically obtained spectra. Finite size effects in our td-MPS calculations are expected to be weak, but quantitative estimates of their size are difficult.}
\label{figGapVibrExct}
\end{figure}

Since $t \gg J$, we can derive the main features of $A_{\rm c}(\nu)$ from a Born-Oppenheimer ansatz where the spinon is assumed to be static. The approximately linear string tension \cite{Grusdt2018PRX} leads to a discrete set of vibrational \cite{Bulaevskii1968} and rotational \cite{Grusdt2018PRX} states in the spectrum. Because rotational excitations have a node in the center, $|\psi_{\Sigma=0}|^2 = 0$, they do not contribute in the expression for $A_c(\nu)$ and are invisible in ARPES. Indications for the lowest vibrational state have been found in various numerical studies \cite{Dagotto1990,Brunner2000,Mishchenko2001}; we provide further evidence in Figs.~\ref{figSummary}, \ref{figGapVibrExct}. At higher energies, the number of string states per unit of energy grows exponentially. In this regime, self-interactions of the string can lead to hybridization and the formation of a broad continuum, which may explain the absence of higher vibrational peaks in the spectrum.

Because the energy gap to the first vibrational string excitation scales as $\Delta_{\rm c} \simeq t^{1/3} J^{2/3}$ \cite{Bulaevskii1968,Brunner2000,Mishchenko2001}, see Fig.~\ref{figGapVibrExct}, the low-frequency regime in Eq.~\eqref{eqConv} is dominated by the spinon spectrum of width $\simeq J \ll \Delta_{\rm c}$:
\begin{equation}
A(\omega,\vec{k}) = A_{\rm s}(\omega - \nu_{\rm c},\vec{k}) Z_{\rm c} \quad \text{for}~\omega \ll \Delta_{\rm c}.
\label{eqAlowE}
\end{equation}
Here $\nu_{\rm c}$ is the ground state energy of the chargon, and $Z_{\rm c}$ denotes the chargon contribution to the quasiparticle weight. Using the Lehmann representation of $A_c(\nu)$ defined below Eq.~\eqref{eqConv}, we see that $Z_{\rm c}$ is related to the ground state string wavefunction by
\begin{equation}
Z_{\rm c} = |\psi_{\Sigma=0}|^2.
\label{eqZcDef}
\end{equation}
It describes the probability for finding geometric strings of length zero: $\Sigma=0$. This chargon contribution $Z_{\rm c}$ depends strongly on the ratio $t$ and the string tension, which is proportional to $J$. Within the FSA and in the considered regime $t \gg J$, the factor $Z_{\rm c}$ contains the only $t$-dependence of the parton spectrum.

The most important consequence of Eq.~\eqref{eqAlowE} is that the entire momentum dependence of the spectrum is captured by the spinon contribution $A_{\rm s}(\omega,\vec{k})$ at strong coupling. We expect that the latter exhibits a quasiparticle structure,
\begin{equation}
A_{\rm s}(\omega,\vec{k}) = Z_{\rm s}(\vec{k}) \delta \l \omega - \omega_{\rm s}(\vec{k}) \r,
\end{equation}
where $Z_{\rm s}(\vec{k})$ denotes the spinon contribution to the quasiparticle residue and $\omega_{\rm s}(\vec{k})$ is the spinon dispersion. Combining the last results, we expect the following structure of the ARPES spectrum at low-energies,
\begin{equation}
A(\omega,\vec{k}) = Z_{\rm c} Z_{\rm s}(\vec{k})  ~\delta \l \omega - \nu_{\rm c} - \omega_{\rm s}(\vec{k}) \r,  ~~\omega \ll \Delta_{\rm c}.
\label{eqAwkUniv}
\end{equation}

In contrast to the chargon properties $Z_{\rm c}$ and $\nu_{\rm c}$, which are universally determined by the geometric strings, the spinon properties $Z_{\rm s}(\vec{k})$ and $\omega_{\rm s}(\vec{k})$ depend on specific properties of the parton model. Hence, ARPES spectra at strong couplings $t \gg J$ in systems with strong local AFM correlations provide direct information about the properties of constituting spinons in the underlying spin model. Such information is usually extracted from studies of the dynamical spin structure factor \cite{Piazza2015,Verresen2018spec,Ferrari2018}, although in that case only pairs of (interacting) spinons can be excited. In the remainder of this paper, we will discuss a microscopic theory constituting spinons in the 2D square lattice Heisenberg model.

A comment is in order about our notion of \emph{constituting spinons}. If the spin system is in a confining phase, as in the case of the 2D Heisenberg AFM with long-range order, isolated spinon excitations cannot exist: there is no spin-charge fractionalization. The strong coupling parton theory above explicitly assumes, however, that the spinon is bound to the chargon. Such mesonic bound states can exist even in a confining phase. In this case the ARPES spectrum is expected to reveal the properties of the constituting spinon, without the strong renormalization effects present e.g. in the spin structure factor due to spinon-spinon interactions. In a possible deconfined phase, free spinon excitations can exist: the constituting spinons are identical to the free spinons in this case. At strong couplings we still expect that spinon-chargon bound states, as described above, will form at low energies. This scenario is realized e.g. in fractionalized Fermi liquids \cite{Senthil2003,Punk2015PNASS}, and in this case the ARPES spectrum of the form in Eq.~\eqref{eqAwkUniv} is expected to directly reveal the properties of free spinons. 

\emph{Imperfections.--}
Our parton description of ARPES spectra above was based on the FSA ansatz and assumed strong coupling, $t \gg J$. Corrections beyond this idealized scenario are expected: For the 2D $t-J_z$ model it has been shown explicitly that the over-completeness of the string basis leads to weak renormalization of the spinon properties by the chargon \cite{Grusdt2018PRX} through Trugman loop processes \cite{Trugman1988}. Similar renormalization is expected to be present in any spin model, but the effect is generically small compared to the string tension \cite{Grusdt2018PRX}. In most models, the latter is of the same order as the spinon bandwidth. We also expect that the dressing of the spinon with the fluctuating geometric string leads to an overall renormalization of the spinon bandwidth, $\omega_{\rm s}(\vec{k}) \to \omega^*_{\rm s}(\vec{k})$. Importantly, at strong couplings such renormalization is independent of the spinon momentum,  $\omega^*_{\rm s}(\vec{k}) = \nu_{\rm FC}  ~\omega_{\rm s}(\vec{k})$ \cite{Grusdt2019PRB}; $\nu_{\rm FC}$ denotes a $\vec{k}$-independent Franck-Condon factor.

When $t$ and $J$ become comparable, the strong coupling ansatz Eq.~\eqref{eqFSA} needs to be modified by including additional correlations. In this case the center-of-mass momentum $\vec{k}$ is shared by the partons, and scattering of the chargon on the spinon is expected to renormalize the bound state dispersion in a $\vec{k}$-dependent way. Moreover, the overall scale of the dispersion is strongly suppressed compared to the bare dispersion of the constituting spinon, by a Franck-Condon factor $\nu_{\rm FC} \ll 1$ \cite{Grusdt2019PRB}.  

Finally, magnon corrections are expected to contribute to the ARPES spectrum. While the initial state $\l \sum_\sigma \c_{\vec{k},\sigma} \r \ket{\Psi_0}$ is expected to have a large overlap with the one-spinon state considered above, it can also contain spinon-plus-magnon (or three-spinon) contributions. The geometric string introduces couplings of the meson-like bound state to collective spin-wave, or magnon, excitations in the system. Together these effects lead to polaronic dressing of the spinon-chargon pair, which is expected to reduce the quasiparticle residue $Z_{\rm c} Z_{\rm s}(\vec{k}) \to Z_{\rm c} Z_{\rm s}(\vec{k}) Z_{\rm m}(\vec{k})$ and add an incoherent magnon contribution $A_{\rm m}(\omega,\vec{k})$ to the idealized spectral function Eq.~\eqref{eqAwkUniv}. These two effects are related by the sum rule, $Z_{\rm m}(\vec{k}) + \int d\omega A_{\rm m}(\omega,\vec{k}) = 1$ for all $\vec{k}$, which yields an estimate how strongly magnon dressing modifies the parton result. 

\emph{Unbound spinon-chargon pairs.--}
Spin systems in a deconfined phase can also support unbound spinon-chargon pairs. In this case the spectral function also becomes a convolution of a spinon and a chargon (or holon) part. Because the center-of-mass momentum can be distributed arbitrarily between the two partons, the convolution includes both frequency and momentum integrals, 
\begin{equation}
A(\omega,\vec{k}) |_{\rm unbound} = \int d \nu d \vec{\kappa} ~A_{\rm s}(\omega - \nu, \vec{k} - \vec{\kappa}) A_{\rm c}(\nu,\vec{\kappa}).
\label{eqAwkUnbound}
\end{equation}
In the absence of a bound state the quasiparticle residue $Z=0$ vanishes, a hallmark of spin-charge separation \cite{Weng1997}. 

The deconfined scenario is realized for example in the 1D $t-J$ model at strong coupling \cite{Lieb1968,Weng1995,Bannister2000,Giamarchi2003}. There, a similar wavefunction as in Eq.~\eqref{eqFSA} can be used to describe the eigenstates of a single hole \cite{Ogata1990,Kruis2004a}, but the string wavefunctions are extended: $\psi_\Sigma^{\rm FSA}(k_{\rm c}) = e^{- i k_{\rm c} \Sigma} / \sqrt{L}$ where $\Sigma \in \mathbb{Z}$ denotes linear string configurations of length $\ell_\Sigma = |\Sigma|$ and $k_{\rm c}$ is the chargon momentum. The spinon wavefunction in 1D can be accurately modeled by a slave-particle mean-field ansatz for spinons forming a Fermi sea \cite{Weng1995,Bohrdt2018}. Magnon corrections in 1D have also been calculated and shown to be small \cite{Bohrdt2018}.

\begin{figure*}[t!]
\centering
\epsfig{file=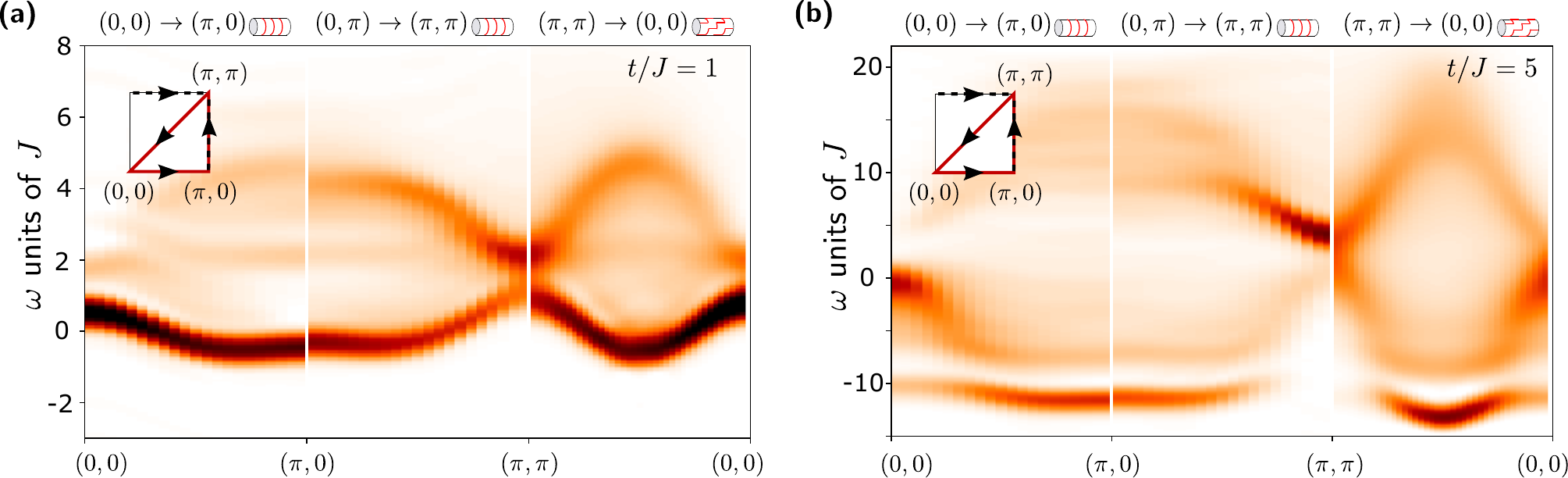, width=0.98\textwidth}
\caption{\textsf{\textbf{One-hole spectral function.}} As in Fig.~\ref{figSummary} of the main text, we use td-MPS methods to calculate the single-hole spectrum in the $t-J$ model on a $4 \times 40$ cylinder. In (a) we used $t/J = 1$ and in (b) we set $t/J=5$.}
\label{figMoreSpectra}
\end{figure*}

\textbf{Numerical results: td-MPS and DMRG.} 
We use a td-MPS method \cite{Zaletel2015} to calculate the ARPES spectrum in the 2D $t-J$ model on a four-leg cylinder, see part A) of the Methods and Ref.~\cite{Bohrdt2019Dyn} for details. In Fig.~\ref{figSummary} (a) the spectrum is shown for $t/J = 3$, well within the strong coupling regime but before the Nagaoka effect plays a role \cite{White2001}. Results for other values of $t/J$, still in the same regime, are qualitatively similar, see Fig.~\ref{figMoreSpectra} in part A) of the Methods section. 

\emph{Ground state and first vibrational excitation.--}
Consistent with earlier spin-wave \cite{Liu1992}, exact diagonalization \cite{Dagotto1990,Leung1995}, truncated basis \cite{Manousakis2007}, cluster-perturbation \cite{Wang2015a} and quantum Monte-Carlo calculations \cite{Brunner2000,Mishchenko2001}, we find a well-defined quasiparticle peak at low energies. The first vibrational peak above the ground state at the nodal point $(\pi/2,\pi/2)$ is also clearly visible, see Fig.~\ref{figSummary} (a). This peak has been found in earlier quantum Monte Carlo studies, which use analytical continuation to obtain the spectral function \cite{Brunner2000,Mishchenko2001}, but the td-MPS method has improved the resolution of the data. 

In Fig.~\ref{figGapVibrExct} we show how the excitation energy $\Delta E$ from ground to the first vibrational state depends on the ratio $t/J$. A linear fit to our MPS data confirms the linear scaling $\Delta E \propto t^{1/3} J^{2/3}$ at strong coupling with remarkable precision. A parameter-free calculation of the excitation gap using the FSA \cite{Grusdt2019PRB} ansatz [as described in part B) of the Methods] is in excellent quantitative agreement with our numerical results.

Owing to the improved resolution of our data, the first vibrational peak is clearly visible in the spectrum for \emph{all} momenta in Fig.~\ref{figSummary} (a) where the ground state quasiparticle weight takes appreciable values. Its energy gap $\Delta E (\vec{k})$ to the ground state is approximately the same for all $\vec{k}$. The dispersion relation of both peaks follows the expected spinon dispersion \cite{Grusdt2019PRB}. 

\emph{Ground state quasiparticle weight.--}
The residue $Z(\vec{k})$ of the ground state quasiparticle peak has a strong momentum dependence, which we attribute to the spinon properties in the parton theory. We find that the spectral weight is strongly suppressed around $(\pi,\pi)$, all the way up to large energies $\simeq 2t$ above the ground state energy. Above this scale, a pronounced dispersive feature is revealed. Outside the magnetic Brillouin zone (MBZ, defined by $|k_x| + |k_y| \leq \pi$) we observe a drop of the quasiparticle residue. Later in this article we will argue that this is a direct signature for fermionic spinon statistics. 

Around $(0,0)$ we also find a suppressed quasiparticle weight, $Z(0,0) < Z(\pi/2,\pi/2)$, but compared to the situation at $(\pi,\pi)$ the effect is less pronounced. More significantly, we find spectral weight in a broad continuum starting slightly above the ground state at $(0,0)$. This should be contrasted with the complete suppression of spectral weight over a wide energy range at $(\pi,\pi)$. In addition, the pronounced high-energy feature at $(\pi,\pi)$ is completely absent at $(0,0)$. These findings indicate that different mechanisms are responsible for the reduction of spectral weight around $(0,0)$ and $(\pi,\pi)$. This finding is further supported by the observation of different $t/J$-dependence at $(0,0)$ and $(\pi,\pi)$, see Fig.~\ref{figMoreSpectra} in part A) of the Methods section.

\begin{figure}[t!]
\centering
\epsfig{file=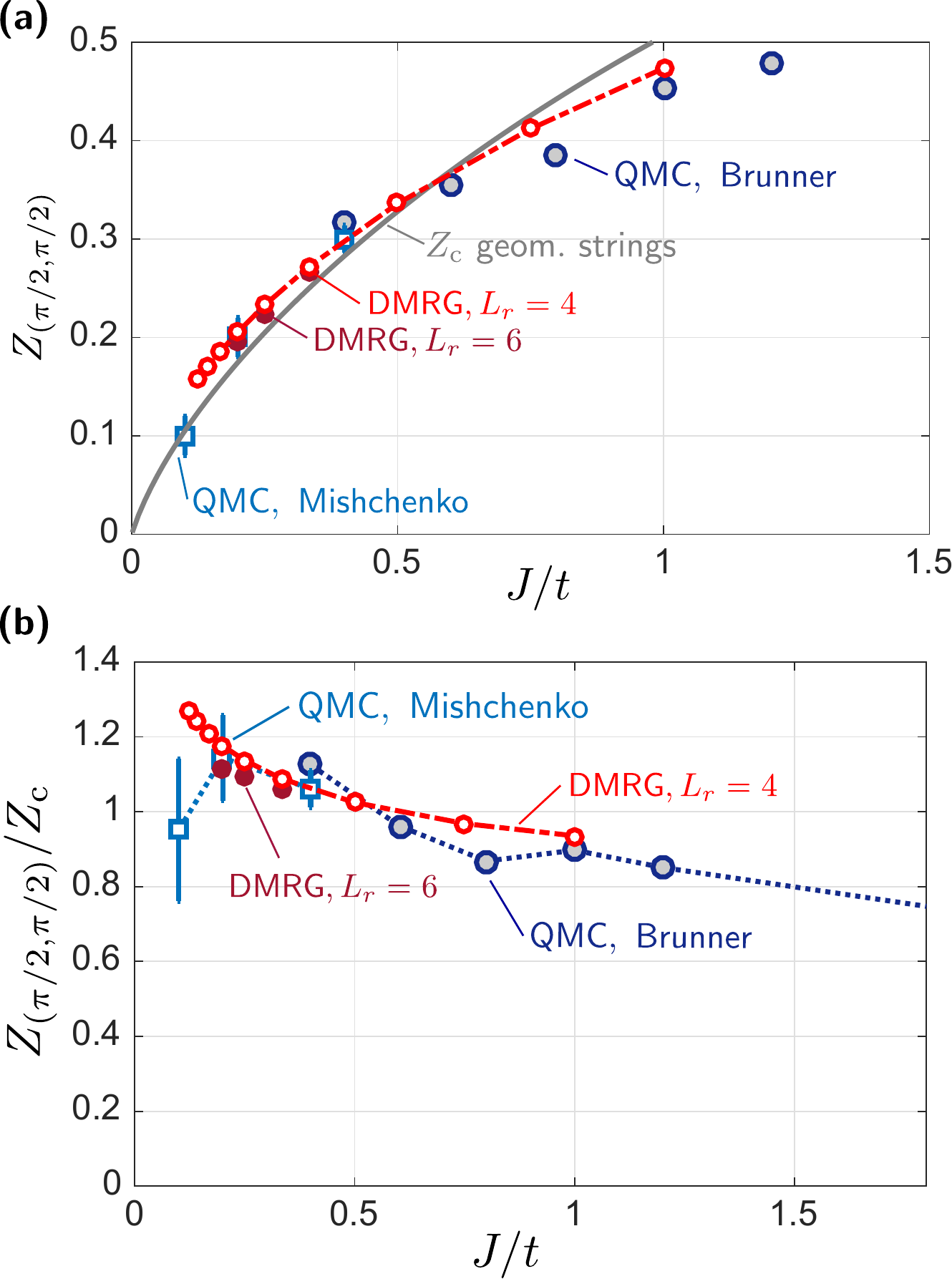, width=0.44\textwidth} $\quad$
\caption{\textsf{\textbf{Ground state quasiparticle weight and parton contributions.}} (a) The quasiparticle weight $Z(\pi/2,\pi/2)$ at the nodal point is shown as a function of $J/t$. We find that earlier numerical Monte Carlo studies by Brunner et al.~\cite{Brunner2000} and Mishchenko et al.~\cite{Mishchenko2001} predict values close to the bare chargon, or string, contribution $Z_{\rm c} = |\psi^{\rm FSA}_{\Sigma=0}|^2$ expected from the geometric string approach (solid gray line). This is confirmed by our DMRG simulations on cylinders with $L_r$ legs -- we used bond dimensions $\chi=500$ ($\chi=600$) for $L_r=4$ ($L_r=6$). (b) We plot $Z_{(\pi/2,\pi/2)} / Z_{\rm c}$ as a function of $J/t$. The data shows only weak dependence on $J/t$, indicating that $Z_{\rm c}(J/t)$ captures the main $J/t$-dependence of the quasiparticle weight.}
\label{figDeptJ}
\end{figure}

In Fig.~\ref{figDeptJ} we show how the ground state quasiparticle weight $Z_{(\pi/2,\pi/2)}(J/t)$ at the nodal point depends on the ratio $J/t$. We compare our DMRG results to earlier quantum Monte Carlo calculations \cite{Brunner2000,Mishchenko2001}. Using DMRG \cite{White1993} we calculate this quantity from the ground state wavefunction, see part A) in the Methods. From the parton theory, we expect that $Z \approx Z_{\rm c} Z_{\rm s}$ factorizes into chargon, or string, and spinon contributions, $Z_{\rm c}$ and $Z_{\rm s}$ respectively, see Eq.~\eqref{eqAwkUniv}. We also argued that, at strong couplings $t \gg J$, only $Z_{\rm c}$ depends on the ratio $J/t$ while $Z_{\rm s}(\vec{k})$ only depends on momentum. Now we check this prediction of the parton theory.

In Fig.~\ref{figDeptJ} (a) we compare the numerical results directly to $Z_{\rm c}(J/t) = |\psi_{\Sigma=0}^{\rm FSA}(J/t)|^2$ which we calculate from the FSA, see Eq.~\eqref{eqZcDef}. Without any free fit parameters, we find that the FSA approach captures correctly the observed $J/t$ dependence, even on a quantitative level. This indicates $Z_{\rm s}(\pi/2,\pi/2) \approx 1$ and additional magnon corrections can be ignored at the nodal point. In Fig.~\ref{figDeptJ} (b) this observation is confirmed by plotting $Z_{(\pi/2,\pi/2)} / Z_{\rm c}$ as a function of $J/t$. There we find that $Z_{(\pi/2,\pi/2)} / Z_{\rm c} \approx 1$ approaches $1$ when $t \gg J$. The dependence of $Z_{\rm c}(J/t)$ on $J/t$ is discussed in part B) of the Methods.

Except for the features at very high energy $\simeq 2 t$ above the ground state, we conclude that the ARPES spectrum at strong coupling can be understood from the general parton theory. In the following we will focus on the quasiparticle weight $Z(\vec{k})$ of the magnetic polaron ground state. We will describe a microscopic theory of spinons, chargons and geometric strings and show that it captures the main features of $Z(\vec{k})$ observed numerically. 

\textbf{Trial wavefunction.} 
The parton approach can be put on a more solid footing by considering a microscopic trial wavefunction describing spinon-chargon pairs in the 2D $t-J$ model. We will demonstrate below that its qualitative predictions are in excellent agreement with the numerical results. Some quasiparticle properties predicted by the trial state are rather sensitive to the variational parameters in the wavefunction however [see part C) in the Methods], which complicates quantitative predictions for $Z(\vec{k})$ or the variationally optimal average string length \cite{Grusdt2019PRB}. 

At quasi-momentum $\vec{k}$, the trial wavefunction we use describes magnetic polarons with fermionic constituting spinons as
\begin{multline}
\ket{\Psi_{\rm sc}(\vec{k})} = \sum_{\vec{j}^s}  \frac{ (u_{\vec{k},\sigma,-}^{(\vec{j}^s)})^* e^{i \vec{k} \cdot \vec{j}^s}}{L / \sqrt{2}} \\
 \times  \sum_\Sigma \psi_\Sigma ~ \hat{G}_\Sigma ~ \hat{\mathcal{P}}_{\rm GW} ~ \f_{\vec{j}^s,\sigma} \ket{\Psi_{\rm MF}^{\rm SF+N}}.
\label{eqDefMesonWvfct}
\end{multline}
We dropped $\hd_{\vec{j}}$ because the state of the single chargon is fully determined by the Gutzwiller projection; $u_{\vec{k},\sigma,-}^{(\vec{j}^s)}$ denotes the cell-periodic part of a Bloch wavefunction. This ansatz is based on a mean-field model of the Heisenberg AFM with constituting fermionic $U(1)$ Dirac spinons $\f_{\vec{j}^{\rm s},\sigma}$, which has attracted renewed interest recently \cite{Song2019}. The mean-field state 
\begin{equation}
\ket{\Psi_{\rm MF}^{\rm SF+N}} = \prod_{\vec{k} \in {\rm MBZ}} \prod_{\sigma} \fd_{\vec{k},\sigma,-} \ket{0}
\end{equation}
is a fermionic band insulator \cite{Baskaran1987}, where $\fd_{\vec{k},\sigma,\nu}$ creates a spinon with band index $\nu = \pm$. The mean-field spinon dispersion $\omega_{\rm s}(\vec{k})$ has been determined variationally  \cite{Lee1988,Piazza2015} to be well described by a model with staggered Peierls flux $\pm \Phi$ per plaquette and a staggered Zeeman field $\pm B_{\rm st} /2$ breaking the $SU(2)$ symmetry \footnote{While the staggered magnetic field $B_{\rm st}$ improves the ground state energy for the half-filled Heisenberg AFM, it has been shown in \cite{Piazza2015} to cause an unphysical gap in the dynamical spin structure factor at momentum $(\pi,\pi)$. The dynamical spin structure factor is a response involving a pair of two constituting spinons, which requires the inclusion of their mutual interactions. In the present work we consider the one-spinon sector, where spinon-spinon interactions play no role. In our case, the staggered field does not lead to inconsistencies.},
\begin{multline}
\H_{f, {\rm MF}} =  - J_{\rm eff} \sum_{\langle \vec{i}, \vec{j} \rangle, \sigma} \l  e^{i \theta^{\Phi}_{\vec{i},\vec{j}}} \fd_{\vec{j},\sigma} \f_{\vec{i},\sigma} + \hc \r \\
+ \frac{B_{\rm st}}{2} \sum_{\vec{j}, \sigma} (-1)^{j_x+j_y} \fd_{\vec{j},\sigma} (-1)^\sigma \f_{\vec{j},\sigma}.
\label{eqDefHfMF}
\end{multline}
A trial state related to Eq.~\eqref{eqDefMesonWvfct}, but without the geometric strings, has been proposed in Ref.~\cite{Giamarchi1993}.

For the square lattice Heisenberg AFM with nearest-neighbor (NN) couplings, the optimized variational parameters are $B^{\rm opt}_{\rm st} / J_{\rm eff} = 0.44$ and $\Phi^{\rm opt} = 0.4 \pi$ \cite{Piazza2015}. In the $t-J_z$ model \cite{Chernyshev1999}, $B^{\rm opt}_{\rm st} / J_{\rm eff} \to \infty$ and the trial wavefunction is highly accurate \cite{Grusdt2018PRX}. In the latter case, the Gutzwiller projection becomes obsolet in Eq.~\eqref{eqDefMesonWvfct} because the mean-field state localizes each spin species on a separate sublattice at half filling.

The  $\vec{k}$-dependent physical properties of the trial wavefunction can be calculated using variational Monte Carlo sampling \cite{Gros1989}. Here we apply this method to calculate the quasiparticle weight [recall that we dropped $\h_{\vec{j}}$ in Eq.~\eqref{eqDefMesonWvfct} above],
\begin{equation}
Z(\vec{k}) = \frac{\sum_\sigma |\bra{   \Psi_{\rm sc}(\vec{k}) }  \hat{f}_{\vec{k},\sigma}  \hat{\mathcal{P}}_{\rm GW} \ket{ \Psi_{\rm MF}^{\rm SF+N} } |^2 }{| \bra{\Psi_{\rm MF}^{\rm SF+N}} \hat{\mathcal{P}}_{\rm GW}  \ket{\Psi_{\rm MF}^{\rm SF+N}}  \bra{\Psi_{\rm sc}(\vec{k})} \Psi_{\rm sc}(\vec{k}) \rangle |},
\label{eqDefZsc}
\end{equation}
where the denominator guarantees proper normalization. Our Monte Carlo procedure for sampling \eqref{eqDefZsc} is explained in part C) of the Methods.

\begin{figure}[t!]
\centering
\epsfig{file=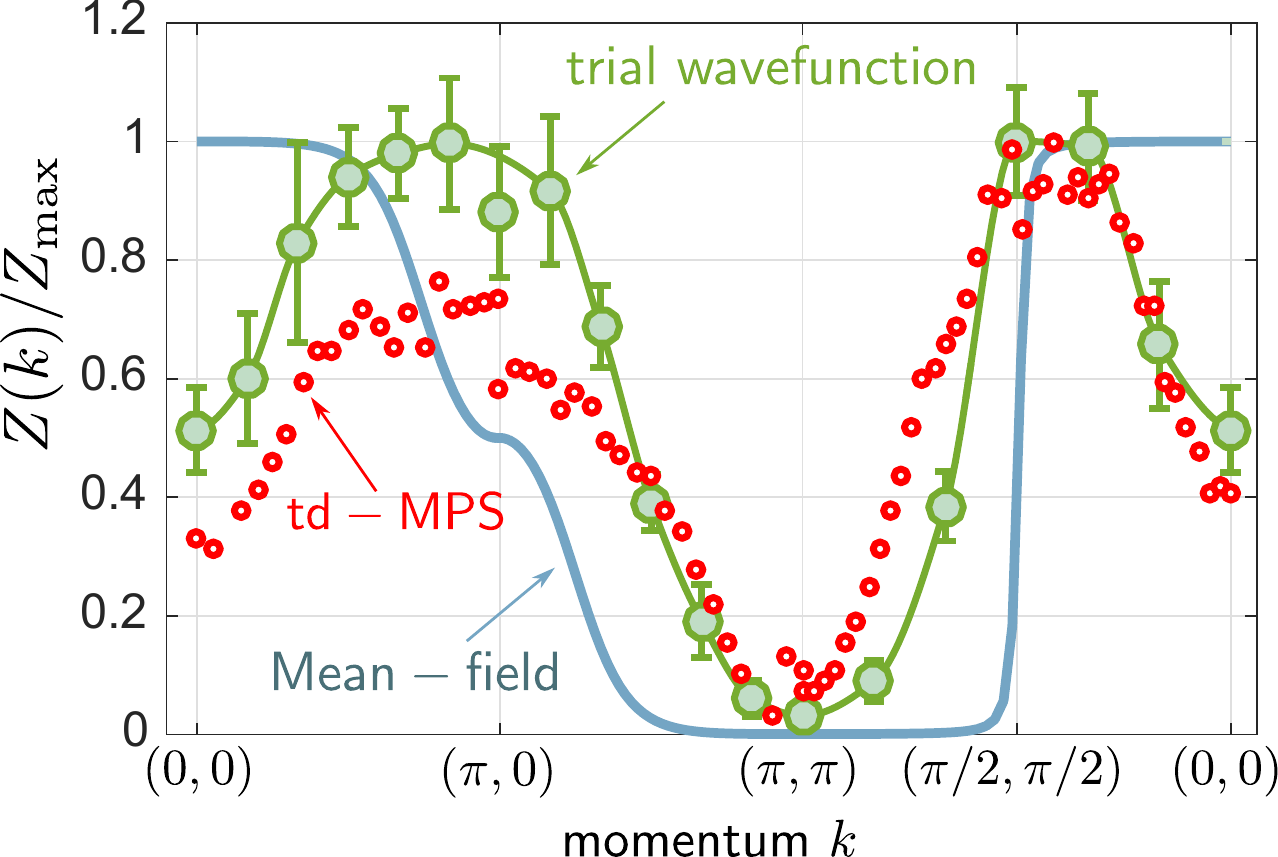, width=0.44\textwidth} $\quad$
\caption{\textsf{\textbf{Momentum dependence of the quasiparticle weight.}} We calculate the quasiparticle residue $Z(\vec{k})$, normalized by $Z_{\rm max} = \max_{\vec{k}} Z(\vec{k})$, from the trial wavefunction \eqref{eqDefMesonWvfct} along a cut $(0,0) - (\pi,0) - (\pi,\pi) - (0,0)$ in a periodic $12 \times 12$ system. Parameters are $t=3 J$ and we used the optimized mean-field parameters $B_{\rm st} / J_{\rm eff} = 0.44$ and $\Phi = 0.4 \pi$. The solid green line is a guide to the eye. We compare our results to the bare mean-field prediction (solid blue line) and results from our td-MPS calculations (red dots). See part A) of the Methods how $Z(\vec{k})$ is numerically extracted from td-MPS.}
\label{figCutK}
\end{figure}

\emph{Results.--}
In Fig.~\ref{figCutK} we plot the $\vec{k}$-dependence of the quasiparticle weight $Z$ of the trial wavefunction. We set $t=3J$, in the strong coupling regime, and used the string wavefunction $\psi_\Sigma = \psi_\Sigma^{\rm FSA}$ obtained from the FSA in Eq.~\eqref{eqDefMesonWvfct}. We checked that no significant dependence on system size remains, see Methods part C). 

The result is in very good agreement with our numerical td-MPS and previous Monte-Carlo results \cite{Brunner2000}. In the center of the MBZ, around $(0,0)$, we observe a dip of the spectral weight. The maximum is found at the edge of the MBZ, including at the high-symmetry points $(0,\pi)$ and $(\pi/2,\pi/2)$. Outside the MBZ, the $Z$-factor is strongly suppressed: at $(\pi,\pi)$ we calculate that it drops below $10^{-2}$. Overall we observe a strong momentum dependence of the $Z$-factor, which is qualitatively captured by the trial wavefunction. The latter includes strong $\vec{k}$-dependence as a consequence of the Fermi statistics that determine the spinon properties in the trial state.

\begin{figure}[t!]
\centering
\epsfig{file=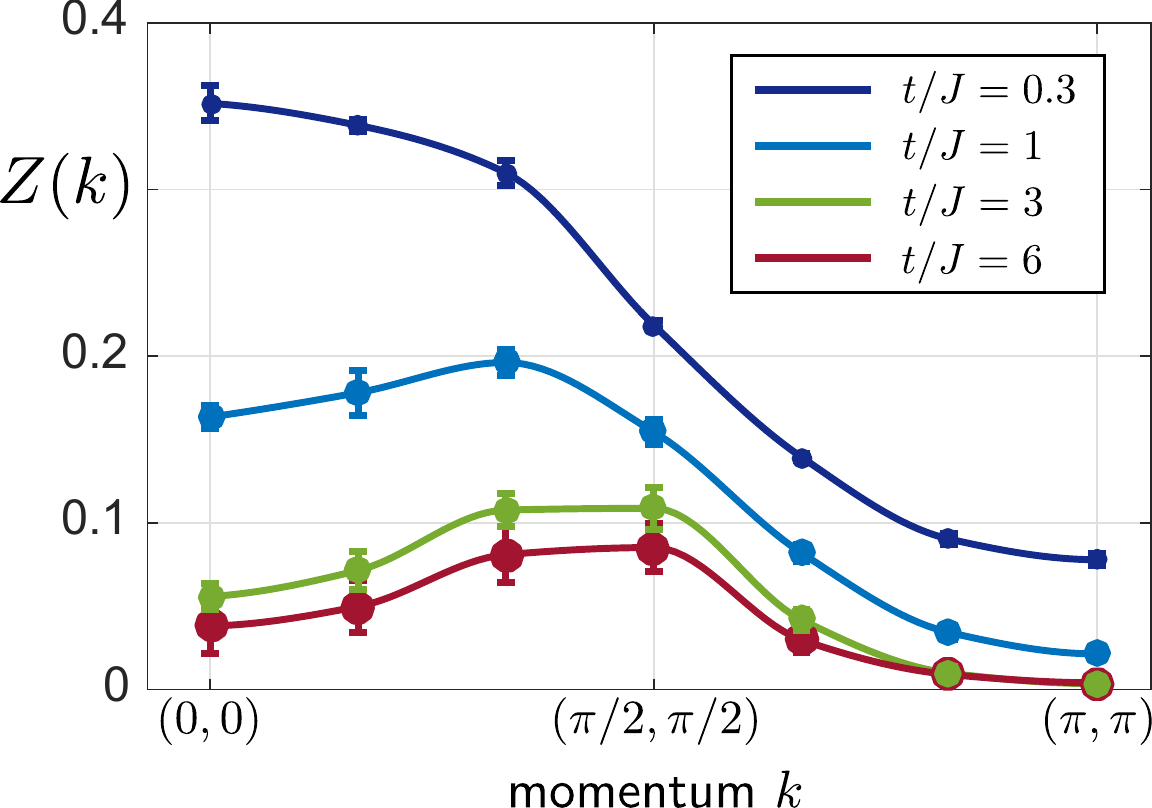, width=0.42\textwidth} $\qquad$
\caption{\textsf{\textbf{Dependence on $t/J$.}} Using the trial wavefunction Eq.~\eqref{eqDefMesonWvfct} with $\psi_\Sigma = \psi_\Sigma^{\rm FSA}$ we calculate the quasiparticle weight $Z(\vec{k})$ along the diagonal cut $(0,0) - (\pi,\pi)$ for different values of $t/J$. We set $B_{\rm st} / J_{\rm eff} = 0.44$ and $\Phi = 0.4 \pi$ in a $12 \times 12$ system; solid lines are guides to the eye.}
\label{figDeptJvmc}
\end{figure}

In Fig.~\ref{figDeptJvmc} we study the dependence of the quasiparticle weight $Z(\vec{k})$ on $t/J$, along the diagonal cut $(0,0) - (\pi,\pi)$. Here $t/J$ controls the length distribution of the geometric strings $\Sigma$ which we use in the FSA string wavefunction $\psi_\Sigma = \psi_\Sigma^{\rm FSA}$ in Eq.~\eqref{eqDefMesonWvfct}. Overall, the spectral weight decreases when $t/J$ is increased, as expected from Eqs.~\eqref{eqAlowE}, \eqref{eqZcDef} on general grounds. In addition, a non-trivial momentum dependence develops as $t/J$ is increased: For small $t/J$ the spectral weight around $(0,0)$ is enhanced, but it is suppressed for large $t/J$. The same qualitative behavior is observed in our td-MPS simulations, see Fig.~\ref{figMoreSpectra} in part A) of the methods. 

The $J/t$ dependence of $Z$ observed in Fig.~\ref{figDeptJvmc} at the nodal point $(\pi/2,\pi/2)$ is also significant. As expected, we observe a decrease in $Z_{(\pi/2,\pi/2)}$ as $t/J$ increases. In comparison with the data shown in Fig.~\ref{figDeptJ}, the overall magnitude of the quasiparticle weight in the trial wavefunction is too small by a factor of about two. We note, however, that the quasiparticle residue depends much more sensitively on the parameters in the trial wavefunction than, for example, the corresponding variational energy \cite{Grusdt2019PRB}. As discussed in more detail in part C) of the Methods, the quasiparticle weight is sensitive to the staggered field $B_{\rm st} / J_{\rm eff}$ and the string tension used to calculate $\psi_\Sigma$ in Eq.~\eqref{eqDefMesonWvfct}. We believe this is the main reason for the observed deviations.

\textbf{Mean-field approximation.}
A mean-field description of the constituting spinons is obtained by dropping the Gutzwiller projection in Eqs.~\eqref{eqDefMesonWvfct}, \eqref{eqDefZsc} and working directly with the mean-field Hamiltonian from Eq.~\eqref{eqDefHfMF}. In principle bosonic or fermionic spinons can both be considered. However, the bosonic theory would require strong interactions to explain the observed spinon quasiparticle weight $Z_{\rm s}(\vec{k}) \approx Z(\vec{k}) / Z_{\rm c}$. We will argue that non-interacting fermionic spinons readily predict the qualitative features of $Z_{\rm s}(\vec{k})$.

\emph{Fermionic spinons.--}
We calculate $Z^{\rm MF}(\vec{k})$ by applying the FSA and mean-field approximations in Eq.~\eqref{eqDefZsc}. First we note that $\vec{k}$ is an arbitrary vector from the full Brillouin zone (BZ); spinon operators $\hat{f}_{\vec{k},\sigma}$ are defined in the BZ, whereas for spinons $\hat{f}_{\vec{k},\sigma,\nu}$ with band indices $\nu$ the cases $\vec{k} \in {\rm MBZ}$ and $\vec{k} \notin {\rm MBZ}$ have to be distinguished. 

In the FSA we assume that only the trivial string state $\Sigma = 0$ contributes, since non-trivial string states are approximately orthogonal to the background AFM. Then, dropping the Gutzwiller projections yields 
\begin{equation}
Z^{\rm MF}(\vec{k}) = Z_{\rm c} \sum_\sigma | \bra{ \Psi_{\rm MF}^{\rm SF+N} } \fd_{\vec{k},\sigma,-}  \f_{\vec{k},\sigma}| \Psi_{\rm MF}^{\rm SF+N} \rangle|^2.
\label{eqZMF}
\end{equation}
This expression is of the general form expected from the parton theory, see Eq.~\eqref{eqAwkUniv}. The spinon contribution on the right hand side of Eq.~\eqref{eqZMF} is related to the mean-field Bloch wavefunction $u_{\vec{k},\sigma,-}^{(A,B)}$ for sites $\vec{j}$ from the $A$, $B$ sublattice respectively [see part D) in the Methods]
\begin{equation}
Z_{\rm s}^{\rm MF}(\vec{k}) = \frac{1}{2} \sum_\sigma \begin{cases}
| u_{\vec{k},\sigma,-}^{(A)} + u_{\vec{k},\sigma,-}^{(B)} |^2, \quad \vec{k} \in {\rm MBZ}\\
| u_{\vec{k},\sigma,-}^{(A)} - u_{\vec{k},\sigma,-}^{(B)} |^2, \quad {\rm else}.
\end{cases} 
\label{eqZsMF}
\end{equation}

One important conclusion is that $Z_{\rm s}^{\rm MF}(\vec{k})$ generally reflects the $\vec{k}$-dependence of the Bloch wavefunctions, which is determined by the parameters $B_{\rm st} / J_{\rm eff}$ and $\Phi$ in the mean-field Hamiltonian Eq.~\eqref{eqDefHfMF}. Moreover, momenta within the MBZ and outside of it are treated separately, causing constructive and destructive interference of the Bloch wavefunctions respectively.  

\begin{figure}[t!]
\centering
\epsfig{file=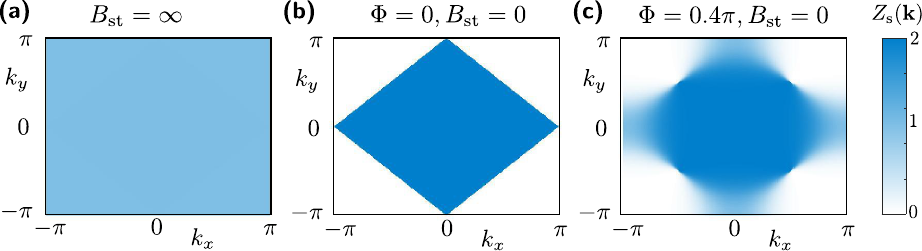, width=0.5\textwidth}
\caption{\textsf{\textbf{Mean-field spinon contribution to the quasiparticle residue.}} $Z_{\rm s}^{\rm MF}(\vec{k})$ from Eq.~\eqref{eqZsMF} is shown in the following limiting cases: (a) $B_{\rm st} / J_{\rm eff} \to \infty$, (b) $B_{\rm st}=\Phi=0$, (c) $B_{\rm st}=0$ and $\Phi=0.4 \pi$. The color bar is indicated on the right.}
\label{figMFZks}
\end{figure}

In the following limiting cases the mean-field spinon contribution $Z_{\rm s}^{\rm MF}(\vec{k})$ shows interesting behavior. In a classical N\'eel state, when $B_{\rm st} / J_{\rm eff} \to \infty$, it holds $(u_{\vec{k},\sigma,-}^{(A)}, u_{\vec{k},\sigma,-}^{(B)}) = (1,0)$ or $(0,1)$. This leads to a featureless spinon spectrum, $Z_{\rm s}^{\rm MF}(\vec{k}) = 1$ everywhere, see Fig.~\ref{figMFZks} (a). For the uniform resonating valence bond state, $\Phi=0$ and $B_{\rm st}=0$, it holds $u_{\vec{k},\sigma,-}^{(A)} = u_{\vec{k},\sigma,-}^{(B)} = 1/\sqrt{2}$. The constituting spinons form a Fermi sea occupying the MBZ, which is directly reflected by the strongly asymmetric spectral weight: $Z_{\rm s}^{\rm MF}(\vec{k}) = 2$ for $\vec{k}$ within MBZ, and $Z_{\rm s}^{\rm MF}=0$ otherwise, see Fig.~\ref{figMFZks} (b). When $B_{\rm st}=0$ but the staggered magnetic flux $\Phi \neq 0$, the mean-field dispersion has a Dirac cone around the nodal point $\vec{k}=(\pi/2,\pi/2)$. This leads to a shard drop of spectral weight along the diagonal from $(0,0)$ to $(\pi,\pi)$ crossing the Dirac point, see Fig.~\ref{figMFZks} (c).

\begin{figure*}[t!]
\centering
\epsfig{file=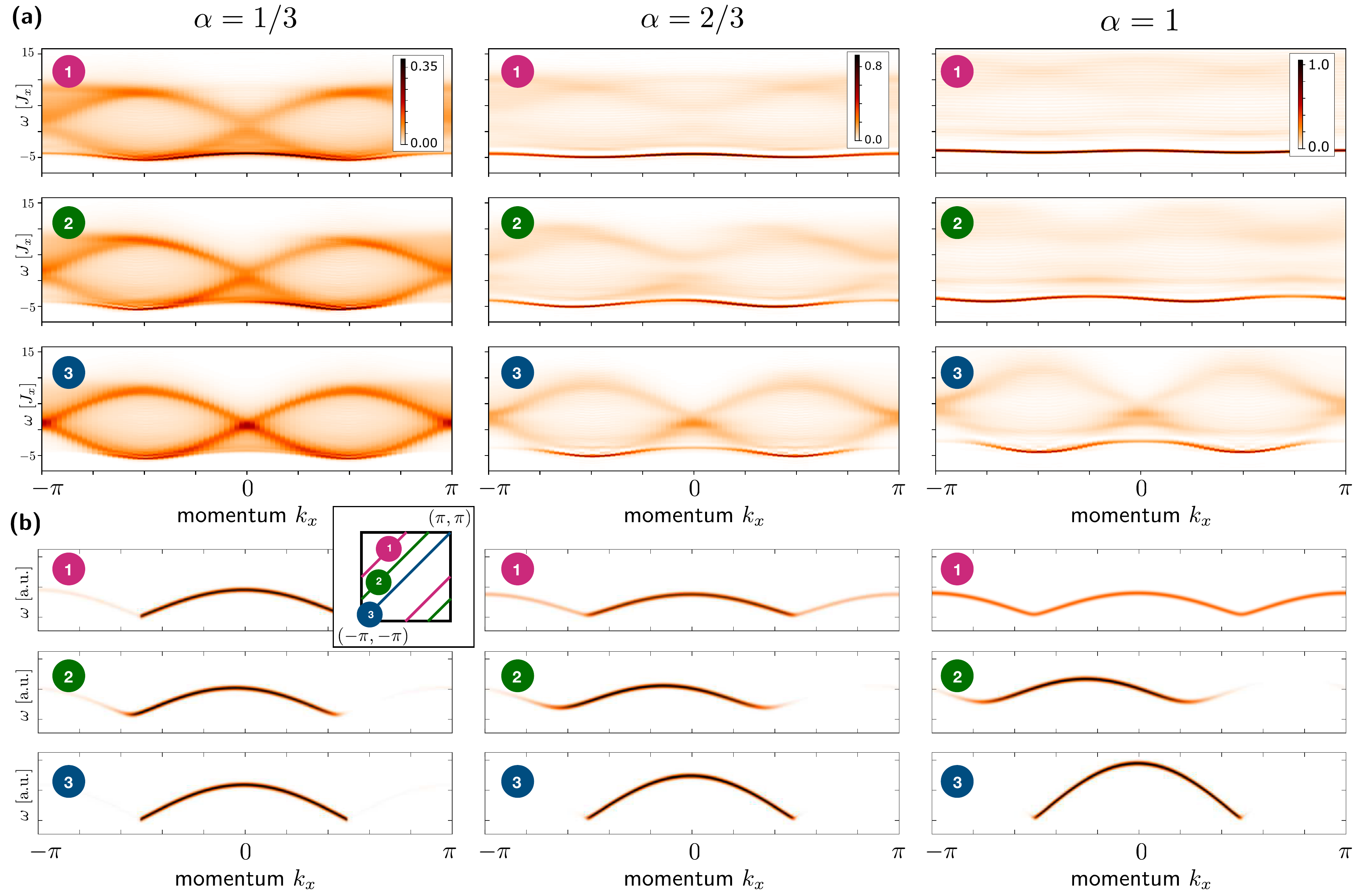, width=0.99\textwidth}
\caption{\textsf{\textbf{One-hole spectral function in the dimensional crossover.}} (a) For varying anisotropy $\alpha = t_y / t_x$ (indicated in top row) of the hopping elements, and $\alpha^2 = J_y / J_x$ of the super-exchange couplings, we use td-MPS to calculate the spectral function. We consider cylinders of length $L_x=40$ along $x$, with circumference $L_y = 4$ along the periodic $y$ direction; $t_x/J_x=3$ is fixed. The MPS is wrapped around the cylinder along diagonals, which allows us to calculate diagonal cuts: $k_y = k_x + k_y^{(0)}$ with $k_y^{(0)} = \pi$, $\pi/2$, $0$ (cuts $1,2,3$ -- see inset below left panel). (b) Predictions for the spinon contribution $Z_{\rm s}$ to the spectral weight (color map) and dispersion from fermionic mean-field theory of spinons, as described in the text. Mean-field parameters are taken from Ref.~\cite{Grusdt2018SciPost}. The delta-function peaks are represented by broadened lines with integrated weight equal to $Z_{\rm s}(\vec{k})$.}
\label{figZcrossover}
\end{figure*}

For the mean-field parameters $B_{\rm st}/J_{\rm eff}$ and $\Phi$ optimized for the half-filled Heisenberg AFM \cite{Piazza2015} the mean-field spinon spectral weight is plotted in Fig.~\ref{figSummary} (c) (color scale) and in Fig.~\ref{figCutK}. We observe a sharp drop of $Z_{\rm s}^{\rm MF}(\vec{k})$ around the nodal point, although the weak staggered field leads to some broadening. Around $(0,\pi)$ and $(\pi,0)$, the decrease of the spectral weight is smoother, which we attribute to the larger distance in $\vec{k}$-space from the Dirac cone found at the nodal point for $B_{\rm st}=0$. 

Overall, the $\vec{k}$-dependence of the quasiparticle weight from the mean-field parton theory, $Z^{\rm MF}(\vec{k}) = Z_{\rm c} Z_{\rm s}^{\rm MF}(\vec{k})$, captures the numerical observations. In particular, it explains the strong suppression of spectral weight around $(\pi,\pi)$, extending up to high energies, as a direct signature of fermionic spinon statistics. Other features observed numerically, such as the more pronounced broadening of spectral weight around the edge of the MBZ and the suppressed quasiparticle residue at $(0,0)$ can be attributed to the Gutzwiller projection and effects beyond FSA. As shown above, these features are correctly predicted by the trial wavefunction Eq.~\eqref{eqDefMesonWvfct}.

\emph{Bosonic spinons.--}
So far we only described the case where the constituting spinons have fermionic statistics. However, using Schwinger bosons, the Heisenberg AFM can also be described by bosonic constituting spinons, see e.g. Ref.~\cite{Auerbach1998}. As described in the Methods [part D)], the trial wavefunction from Eq.~\eqref{eqDefMesonWvfct} can be adapted to the bosonic case. On the mean-field level one finds that the spinon contribution to the quasiparticle residue contains two delta-function peaks of equal weight around $\vec{k} = (0,0)$ and $(\pi,\pi)$. This observation is inconsistent with unbiased numerical results, where $Z(\vec{k})$ is strongly suppressed around $\vec{k} = (\pi,\pi)$. Thus our results favor parton theories with fermionic constituting spinons, although it is difficult to rule out scenarios with bosonic spinons and strong interaction effects suppressing $Z_{(\pi,\pi)}$.

\textbf{Dimensional cross-over.}
Now we subject the parton theory to another test and study ARPES spectra in the dimensional cross-over. We tune the ratio
\begin{equation}
\alpha = t_y / t_x
\end{equation}
of tunneling amplitudes along $x$ and $y$ directions, which leads to spin-exchange couplings $J_y = \alpha^2 J_x$. In ultracold atom experiments with optical lattices, the value of $\alpha$ can be easily tuned. 

Our main motivation for considering the dimensional cross-over is that the parton theory with fermionic spinons correctly describes the ARPES spectrum in the 1D $t-J$ model \cite{Weng1995,Bohrdt2018}. For non-zero $\alpha > 0$ we expect a non-vanishing string tension $\propto \alpha J_x$ which should lead to spinon-chargon binding. At $\alpha = 1$ we have  established above that the parton theory can explain the numerically observed spectra.

In the 1D case, spinons and chargons are deconfined and unbound for $t \gg J$. Because the spectrum is a convolution of spinon and chargon contributions both in momentum and frequency domains, see Eq.~\eqref{eqAwkUnbound}, a coherent quasiparticle peak is absent. Nevertheless, the integrated spectral weight $Z_{\omega}(k) = \int_0^{\Delta \omega} d \nu ~ A(\nu_0(k) + \nu, k)$ in a low-energy region of width $\Delta \omega = \mathcal{O}(J)$ around the ground state at $\nu_0(k)$ reveals the structure of the spinon Fermi sea. Because $t \gg J$, only chargon states from a narrow range of momenta $\Delta k_{\rm c}$ around the minimum at $k_{\rm c} = 0$ of the chargon dispersion $\nu_{\rm c}(k_{\rm c}) = - 2 t \cos (k_{\rm c}) $ contribute to $Z_{\omega}(k)$ \footnote{The chargon and spinon dispersions are only defined up to an overall gauge choice shifting their momenta in opposite directions.}; hence $Z_{\omega}(k) \propto Z_{\rm s}(k)$. In one dimension, the mean-field parton theory for the optimized parameter $B_{\rm st} = 0$ in Eq.~\eqref{eqDefHfMF} predicts a step function, see Eq.~\eqref{eqZsMF},
\begin{equation}
Z_{\rm s}^{\rm MF}(k) = \begin{cases}
2, \quad |k| \leq \pi/2\\
0, \quad {\rm else},
\end{cases} 
\end{equation}
which directly reflects the Fermi-Dirac distribution of spinons in the ground state \cite{Weng1995,Bohrdt2018}. 

In 1D, the above argument predicts a strong suppression of spectral weight up to energies of order $\mathcal{O}(2 t) \gg J$ around $k=\pi$, which has been observed numerically \cite{Szczepanski1990,Bannister2000,Bohrdt2018}. As discussed earlier, we find the same phenomenology around $\vec{k} = (\pi,\pi)$ in two dimensions, where we also attributed the effect to the underlying fermionic spinon statistics in the mean-field parton theory. To further support our argument that the 1D and 2D cases are due to the same physical principle, now we demonstrate that they continuously evolve into each other in a dimensional cross-over. 

In Fig.~\ref{figZcrossover} (a) we show our numerical td-MPS results for values of $\alpha = 1/3, 2/3$ and $1$. We consider three cuts along diagonals, $k_y=k_x+k_y^{(0)}$ with $k_y^{(0)}=\pi,\pi/2$ and $0$. For $\alpha=1/3$ the spectrum still closely resembles the 1D case, and only a weak dependence on $k_y$ is observed: The minima of the ground state dispersion in the second cut, corresponding to $k_y^{(0)} = \pi/2$, are slightly displaced to the left of $k_x = \pm \pi/2$, as expected from the mean-field spinon dispersion. While some spectral weight appears at $\vec{k}=(\pi,0)$ (first cut with $k_y^{(0)} = \pi$), it remains absent over a broad energy range at $\vec{k}=(\pi,\pi)$ (third cut with $k_y^{(0)} = 0$). In general, the high-energy features can still be understood from a theory of quasi-free spinons and chargons as in 1D. 

For $\alpha=2/3$ a well-defined quasiparticle peak is visible at low energies. This is expected from the parton theory, which predicts the formation of a spinon-chargon bound state as soon as the string tension $\propto J_y$ becomes sizable. Around $\vec{k}=(\pi,\pi)$ we still observe a strong suppression of spectral weight over a wide energy range of order $\mathcal{O}(2 t_x)$. The dispersive features at high energies, reminiscent of a free chargon branch, become increasingly less pronounced as $\alpha$ approaches $1$.

In Fig.~\ref{figZcrossover} (b) we plot the mean-field spinon dispersion expected for the dimensional cross-over. While the overall scale is difficult to predict, the shape of the spinon dispersion resembles the numerically observed quasiparticle dispersion for all considered values of $\alpha$. The variational mean-field parameters $B_{\rm st}(\alpha)$ and $\Phi(\alpha)$ have been taken from Ref.~\cite{Grusdt2018SciPost}. The color-scale in Fig.~\ref{figZcrossover} (b) indicates the spinon quasiparticle weight $Z_{\rm s}^{\rm MF}(\vec{k})$. Around the nodal point $\vec{k}=(\pi/2,\pi/2)$ the numerically obtained spectrum, as a function of momentum, evolves significantly more smoothly than expected from the mean-field theory. We attribute this to the effect of the Gutzwiller projection neglected in the mean-field calculation, as discussed earlier for the 2D case. Around $\vec{k}=(\pi,\pi)$ the mean-field theory correctly predicts the strongly suppressed quasiparticle weight at all values of $\alpha$. We conclude that the parton theory correctly predicts the observed qualitative features of the ARPES spectrum in the dimensional cross-over.

\textbf{Frustrated magnetism: Dirac spin liquid scenario.}
The main signature of fermionic spinon statistics we reported so far was indirect and based on the suppression of spectral weight around $\vec{k}=(\pi,\pi)$. Now we use our trial wavefunction and discuss a case where direct signatures for the formation of fermionic Dirac spinons can be observed. These considerations are relevant to doped quantum magnets with additional frustration. Which spin liquid is relevant depends on the type of frustration, and in many cases this is subject of debate. Here, we restrict ourselves to the Dirac spin liquid case captured by our trial wavefunction, while different signatures can be expected for other quantum spin liquids \cite{Wen2004,Savary2017,Knolle2019}.

On the level of the mean-field theory in Eq.~\eqref{eqDefHfMF}, the spin liquid without N\'eel order is realized when $B_{\rm st}=0$ and a non-vanishing staggered flux $\Phi \neq 0$ is considered. This leads to the formation of Dirac cones at the nodal point $\vec{k} = (\pi/2,\pi/2)$, across which the mean-field approach predicts a sudden drop of the spinon quasiparticle weight, see Fig.~\ref{figMFZks} (c). Now we will go beyond the mean-field approximation and demonstrate that the trial wavefunction Eq.~\eqref{eqDefMesonWvfct}, including the Gutzwiller projection, exhibits the same features. 

\begin{figure}[t!]
\centering
\epsfig{file=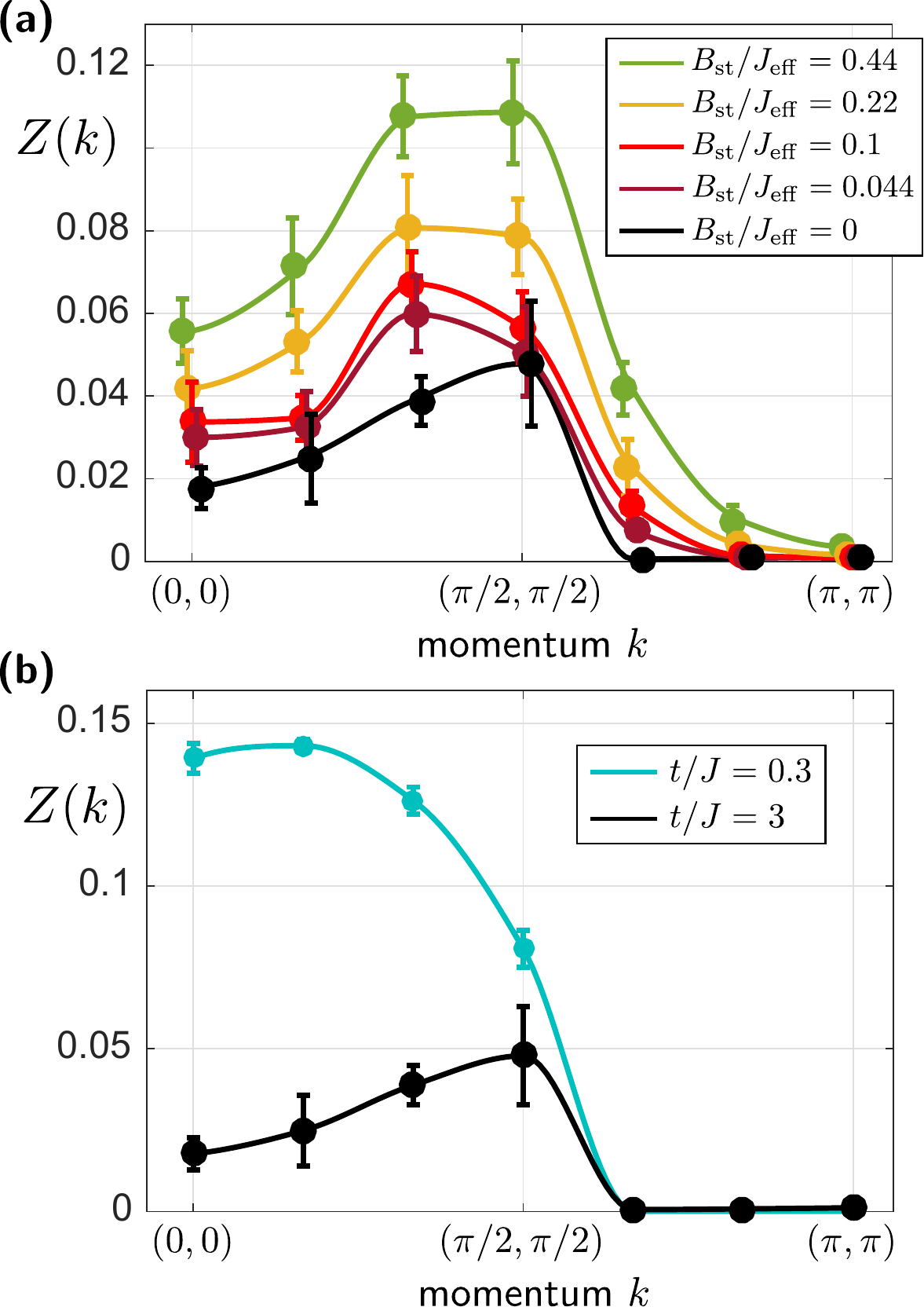, width=0.4\textwidth} $\quad$
\caption{\textsf{\textbf{Quasiparticle weight $Z(k)$ for a Dirac quantum spin liquid scenario}}, along the diagonal cut from $(0,0)$ to $(\pi,\pi)$. (a) When the long-range AFM order in the trial wavefunction, controlled by $B_{\rm st} / J_{\rm eff}$, is reduced, a sudden drop develops around the edge of the MBZ at $(\pi/2,\pi/2)$. Such behavior can be expected if a Dirac quantum spin liquid emerges upon frustrating spin-exchange interactions in the $t-J$ model. This scenario can be described qualitatively by the mean-field theory of fermionic spinons. We used $t/J=3$ and $\Phi=0.4 \pi$ in our trial wavefunction; at each momentum data points are slightly offset horizontally for better visibility. (b) For $B_{\rm st}=0$ and a large range of $t/J$ we observe a sharp drop of spectral weight around the nodal point. In (a) and (b) the solid lines are guides to the eye only.}
\label{figZBstg}
\end{figure}

So far we studied spinon-chargon pairs in a background with long-range N\'eel order, characterized by $B_{\rm st} \neq 0$ in the trial wavefunction. Even in the strong coupling regime, where $t \gg J$, we found that the spectral weight $Z(\vec{k})$ decays smoothly along the diagonal from $\vec{k} = (\pi/2,\pi/2)$ to $(\pi,\pi)$ in 2D, see e.g. Figs.~\ref{figCutK}, \ref{figDeptJvmc}. While a small amount of broadening is expected even on the mean-field level when $B_{\rm st} \neq 0$, most of the effect is due to the Gutzwiller projection in the trial state. 

In Fig.~\ref{figZBstg} (a) we show how the quasiparticle weight $Z(\vec{k})$ of the trial wavefunction depends on the staggered field $B_{\rm st}$. In addition to an overall decrease in magnitude, the formation of a sharp feature at the edge of the MBZ is observed when $B_{\rm st}$ is decreased. For a hole moving in the $SU(2)$-invariant $U(1)$ Dirac spin liquid with $B_{\rm st}=0$ and $\Phi = 0.4 \pi$, we find that $Z(\vec{k}) = 0$ vanishes within errorbars for momenta along the diagonal beyond the nodal point, with $|\vec{k}| > \pi / \sqrt{2}$. This behavior can be understood from the mean-field theory as a direct signature for (i) the formation of a Dirac cone at the nodal point and (ii) fermionic spinon statistics. 

The sudden drop of the quasiparticle weight around the nodal point is reminiscent of the missing spectral weight on the backside of the Fermi arcs observed in the pseudogap phase of cuprates \cite{Shen2005}. Within our microscopic approach, the strong suppression of the quasiparticle residue in the vicinity of the nodal point, but outside the MBZ, is explained by the underlying structure of constituting fermionic spinons. Our calculation demonstrates that the Gutzwiller projection does not necessarily broaden the spectrum in $\vec{k}$-space. As shown in Fig.~\ref{figZBstg} (b), this phenomenology is not necessarily related to the presence of strings in the trial state: it can also be observed for shorter strings when $t < J$. 

The situation described above, with a single hole moving in an $SU(2)$ invariant spin background $B_{\rm st}=0$, is directly relevant to the magnetic polaron problem in the $J_1-J_2$ model on a square lattice. Upon increasing the frustrating diagonal next-nearest neighbor coupling $J_2$, the staggered magnetization in the undoped system approaches zero \cite{Ferrari2018}, corresponding to the choice $B_{\rm st} \to 0$ in the variational wavefunction. We expect that the spinon-chargon trial wavefunction can be used to describe a single mobile hole in the frustrated $J_1-J_2$ background, as long as the tunneling rate dominates, $t \gg J_{1,2}$, and strong local AFM correlations are present.

~ \\
\textsf{\textbf{\large Discussion}}\\
In this article, we discussed a microscopic parton theory of ARPES spectra in 2D quantum AFMs. At strong couplings, where the tunneling rate $t$ dominates over spin-exchange terms $J$, the geometric string approach \cite{Grusdt2018PRX,Chiu2019Science,Grusdt2019PRB} allows us to approximate the spinon-chargon wavefunction as product state of Born-Oppenheimer type. We showed that this results in an ARPES spectrum which can be written as a convolution of a spinon and a chargon (or string) contribution. When the spinon and chargon form a bound state, the chargon contribution provides a $\vec{k}$-independent overall renormalization $Z_{\rm c}$, whereas the spinon contribution $Z_{\rm s}(\vec{k})$ is strongly $\vec{k}$-dependent in general. Conversely, only $Z_{\rm c}(t/J)$ depends on the ratio $t/J$, reflecting the size of the spinon-chargon bound state, whereas the spinon contribution $Z_{\rm s}$ is independent of the tunneling amplitude $t$.

We demonstrated that the ARPES spectrum of a single hole in the 2D $t-J$ model, characterizing the structure of magnetic polarons, can be described by the parton theory. In particular we established that the spinon part of the quasiparticle weight, $Z_{\rm s} = Z / Z_{\rm c}$, becomes only weakly dependent on $t/J$ in the strong coupling limit $t \gg J$ where our theory is valid; $Z_{\rm c}(t/J)$ can be calculated from a semi-analytical string-based calculation. Using td-MPS simulations \cite{Zaletel2015,Bohrdt2019Dyn}, we calculated the momentum dependence of the spectral weight $A(\omega,\vec{k})$, which is strongly suppressed over a wide energy range around $\vec{k} = (\pi,\pi)$. Using the parton theory, we argued that this suppression can be understood as a signature of fermionic spinon statistics. We supported this conclusion by showing that all qualitative features of $Z(\vec{k})$ can be reproduced by a spinon-chargon trial wavefunction based on fermionic constituting spinons and including string-like spin-charge correlations \cite{Grusdt2019PRB}. 

We obtained even more direct signatures for the formation of fermionic spinons for a doped Dirac spin liquid. Across the location of the spinon Dirac cone in the Brillouin zone, we predict a sharp drop of the quasiparticle weight using our trial wavefunction. We explained this feature by a mean-field theory based on constituting fermionic spinons, whose quasiparticle residue changes abruptly across the Dirac cone at the nodal point. 

Our work establishes a possible link to the Fermi arcs observed in the pseudogap phase of cuprates. We suggest that our observation that the spectral weight is strongly suppressed up to high energies around $\vec{k} = (\pi,\pi)$ is a precursor of the missing spectral weight outside the MBZ in the context of Fermi arcs. We identified two important ingredients required for this phenomenology: (i) strong couplings, $t \gg J$, leading to extended geometric strings, and (ii) sufficient frustration, leading to a small AFM order parameter. In the cuprate compounds, typical values of $t/J$ are around $3$, well within the strong coupling regime. Next we will argue that the required kind of frustration is naturally introduced by the mobile dopants themselves in the pseudogap regime.

\textbf{Parton theory at finite doping.} 
The parton theory can be easily extended to finite but small doping, if we assume that correlations between the chargons can be neglected and they remain bound to individual spinons by geometric strings. Experimental studies in ultracold atom systems suggest that these conditions may be justified up to a maximum doping level of about $15 \%$ \cite{Chiu2019Science,Bohrdt2019NatPhys}. Beyond this regime magnetic polarons begin to overlap, and interactions can modify our picture significantly. In the following we focus on the low doping case.

As long as the geometric string picture can be applied, the dispersion of magnetic polarons is dominated by the spinon properties. The number of spinons $N_{\rm s} = L^2 ( 1 -  n_{\rm h})$ decreases with increasing hole doping $n_{\rm h}$, and on the mean-field level the spinon Fermi sea, or band insulator for $B_{\rm st} \neq 0$, is below half filling. Hence fewer states contribute to the ARPES spectrum, and at the lowest energy we expect to see the spinon Fermi surface. 

The variational parameters $B_{\rm st} / J_{\rm eff}$ and $\Phi$ of the mean-field spinon Hamiltonian Eq.~\eqref{eqDefHfMF} are expected to depend on the doping level $n_{\rm h}$. For $n_{\rm h}=0$ the optimal parameters correspond to the half-filled Heisenberg model \cite{Piazza2015}, which we used in this article. At finite doping $n_{\rm h}>0$, in contrast, the geometric strings introduce effective next-nearest neighbor (and further) interactions $J_2 (n_{\rm h})$ in the spin background $\ket{\Psi}$ used to define the FSA product wavefunction Eq.~\eqref{eqFSA}: In a state $\ket{\Sigma} = \hat{G}_\Sigma \ket{\Psi}$, the spins along the geometric string $\Sigma$ are displaced by one site. The instantaneous spin-exchange coupling $\H_J$ in state $\ket{\Sigma}$ hence includes interactions between spins that used to be next-nearest neighbors in the original state $\ket{\Psi_0}$. Averaging over all string configurations $\Sigma$ contributing to the spinon-chargon bound states thus leads to the estimate
\begin{equation}
J_2(n_{\rm h}) \simeq J n_{\rm h} \overline{\ell}_\Sigma,
\end{equation}
where $\overline{\ell}_\Sigma$ is the average length of geometric strings. By the same argument, nearest neighbor interactions $J_1$ are effectively reduced: $J_1(n_{\rm h}) = J (1 - n_{\rm h} \overline{\ell}_\Sigma)$.

As discussed earlier, the presence of frustrating next-nearest neighbor (and further-range) couplings $J_2(n_{\rm h})$ leads to reduced AFM order $B_{\rm st}$ in the mean-field spinon Hamiltonian \cite{Ferrari2018}. Finite temperature is expected to have a similar effect. Once $B_{\rm st}=0$, the trial wavefunction no longer breaks the discrete translational symmetry of the square lattice. In this case, the spinons can still form a small Fermi surface, and the spinon-chargon pairs could form a fractionalized Fermi liquid state \cite{Senthil2003,Punk2015PNASS}. From our insights obtained for a single dopant, we expect that the backside of the corresponding spinon Fermi surface would be invisible in the spectral function, as a consequence of the formation of gapless Dirac cones at the nodal points in the spinon dispersion. The mean-field theory further predicts that the spectral weight smoothly vanishes as one encircles the nodal point, see Fig.~\ref{figMFZks} (c). This phenomenology is in agreement with the experimental findings \cite{Shen2005}.

\textbf{Probing frustrated quantum magnets.} 
Beyond the $t-J$ and, by extension, the Fermi-Hubbard model, our results are also relevant to other strongly correlated quantum spin systems. Our main assumptions within the geometric string approach are (i) that we work in the strong coupling regime, where the tunneling $t$ dominates over spin-exchange terms, and (ii) that the string basis used for the formulation of the FSA wavefunction in Eq.~\eqref{eqFSA} is valid. We expect that (ii) can be satisfied, provided that the system has sufficiently strong \emph{local} AFM correlations.

Under these assumptions, our microscopic parton theory describes general quantum AFMs. According to our results, the corresponding ARPES spectrum should directly reveal the properties of the constituting spinons, including the shape of their dispersion relation and the distribution of spectral weight. Such studies are similar to calculations of the dynamical spin-structure factor, see e.g. Refs.~\cite{Gohlke2017,Ferrari2018,Verresen2018spec}, but as a main advantage they involve one instead of two spinons. This approach may prove to be particularly useful to reveal the nature of quantum spin liquids with deconfined spinon excitations. Concrete examples may include studies of the frustrated Heisenberg model on a Kagome lattice \cite{Depenbrock2012}, where it remains debated wether the ground state is gapped $\mathbb{Z}_2$ or a gapless Dirac spin liquid, or the $J_1-J_2$ model on a triangular lattice where new signatures of Dirac spin liquids have recently been reported \cite{Hu2019}. Our studies are of particular interest in light of the recent proposal that $U(1)$ Dirac spin liquids may provide a unified starting point for describing a range of 2D quantum magnets on different lattices \cite{Song2019}.

\textbf{Experimental considerations.} 
ARPES is a standard tool in solid state physics, and has been used extensively to study strongly correlated quantum matter such as the cuprate compounds \cite{Damascelli2003}. The presence of phonons makes a direct comparison of experimental ARPES spectra and theoretical calculations for the simplified $t-J$ model challenging \cite{Cuk2005,Kar2008}. However, the rapid progress of quantum simulation experiments with ultracold atoms has recently enabled experiments in clean model systems with tunable parameters, where ARPES measurements can also be performed \cite{Kollath2007,Stewart2008,Greif2011,Torma2016,Bohrdt2018,Brown2019}. In particular, the 2D Fermi Hubbard model can be studied and long-standing questions about strongly correlated quantum matter can now be addressed. 

In optical lattices, the lowest temperatures have been achieved in quantum gas microscopy setups so far \cite{Mazurenko2017}, which can also be used to measure the spectral function \cite{Bohrdt2018,Brown2019}. Experiments implementing triangular lattices are currently under construction, paving the way for spectroscopic studies of highly frustrated quantum magnets in the near future. Other applications include systematic investigations of the dimensional cross-over, which we studied in the present paper, or studies of ARPES spectra in bilayer systems.

~
\newpage

~ \\
\textsf{\textbf{\large Methods}}\\
\textsf{\textbf{A) td-MPS simulations and DMRG.}}

\emph{Calculating the spectral function.--}
The spectral function is calculated as the Fourier transform of the time-dependent correlation function
\begin{equation}
C_{\mathbf{i},\mathbf{j}}(t) = \sum_\sigma \bra{\psi_0} e^{i\hat{H}t}\hat{c}_{\mathbf{j},\sigma}^\dagger e^{-i\hat{H}t} \hat{c}_{\mathbf{i},\sigma} \ket{\psi_0}.
\end{equation}
Here $\ket{\psi_0}$ is the ground state of the $t-J$ model without a hole, on a cylinder with four legs. The time evolution of the ground state with the $t-J$ Hamiltonian  $e^{i\hat{H}t} \ket{\psi_0} = e^{iE_0t} \ket{\psi_0}$. We thus calculate
\begin{itemize}
\item the ground state without a hole, $\ket{\psi_0}$, using DMRG
\item the time evolution of the ground state after a hole was created in the origin, $\ket{\psi(t)} = e^{-i\hat{H}t} \hat{c}_{\mathbf{0},\sigma} \ket{\psi_0}$
\item the overlap of $\ket{\psi(t)}$ with the state where a hole was created at a position $\mathbf{j}$, $\ket{\psi_1} = \hat{c}_{\mathbf{j},\sigma} \ket{\psi_0}$.
\end{itemize}
The time evolution of $\hat{c}_{\mathbf{0},\sigma} \ket{\psi_0}$ is performed using the matrix product operator based time evolution introduced in Ref.~\cite{Zaletel2015}; see also Refs.~\cite{Kjall2013,Gohlke2017,Verresen2018spec,Paeckel2019}.

\begin{figure}[t!]
\centering
\epsfig{file=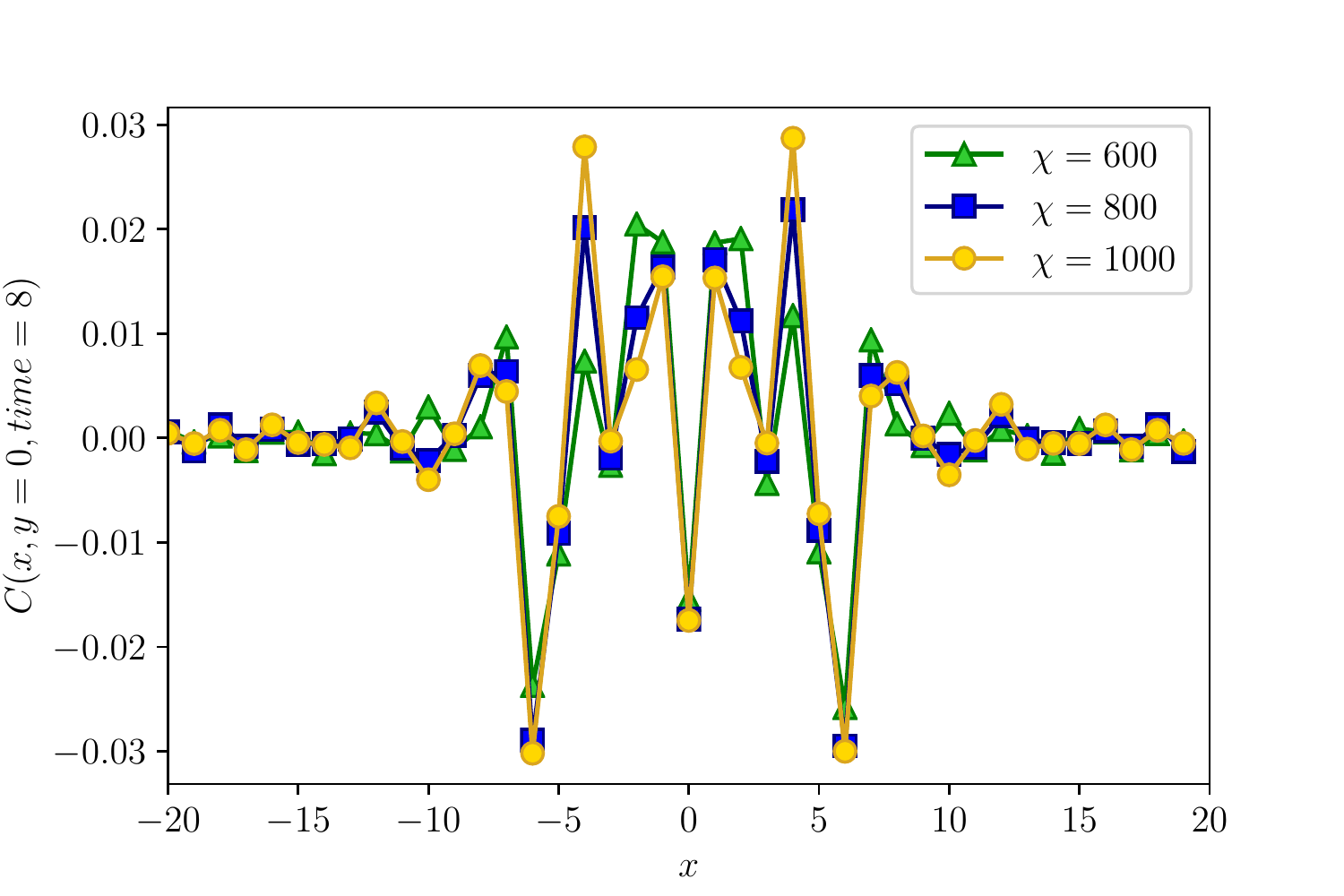, width=0.49\textwidth}
\caption{\textsf{\textbf{Convergence of td-MPS data.}}  We check the convergence of our td-MPS calculations with the bond dimension $\chi$ in the real time and space correlation function $C_{\mathbf{i},\mathbf{j}}(t)$ for the latest time ($8$ in units of $1/J$) considered for $t/J=3$.}
\label{figtdMPSconv}
\end{figure}

In Fig.~\ref{figtdMPSconv} we compare the correlation function $C_{\mathbf{i},\mathbf{j}}(t)$ for different bond dimensions $\chi$ at a time of $t_0 = 8$ ($1/J$), which is the maximal time used for our calculation of the spectral function shown in the main text. While there are small differences in the absolute numbers, the qualitative behavior is captured correctly already at a bond dimension of $\chi=600$. For later times, the deviations between different bond dimensions increase further. 

We perform a spatial Fourier transform to obtain
\begin{equation}
A(\mathbf{k},t) = \sum_{\mathbf{j}} e^{-i\mathbf{k}\cdot \mathbf{j}} C_{\mathbf{0},\mathbf{j}}(t).
\end{equation}
As our time evolution is limited, we use linear prediction to increase the time window. Afterwards, the data is multiplied with a Gaussian envelope~\cite{Verresen2018spec}. Fourier transforming in time yields the spectral function
\begin{equation}
A(\mathbf{k},\omega) = \frac{1}{2\pi} \int_{-\infty}^\infty dt A(\mathbf{k},t).
\end{equation}
In this signal, the Gaussian envelops introduced in the time domain before lead to Gaussian broadening of the obtained peaks.

The diagonal cut through the Brillouin zone from $(0,0)$ to $(\pi,\pi)$ is obtained by labeling the sites around the cylinder in a \emph{zigzag} fashion as indicated in the top row of Fig.~\ref{figSummary} (a). One ring around the cylinder with this labeling contains $2 L_r$ instead of $L_r$ sites, where $L_r$ is the circumference of the cylinder. We calculate the ground state as well as the dynamics with the couplings in the Hamiltonian according to this modified lattice geometry used for representing the MPS. For each time step, we obtain an array with $L_x \times 2 L_r$ entries, where $L_x$ is the length of the cylinder. This array is reshaped into an $2 L_x \times L_r$ array and then the Fourier transform is performed, yielding $A(\mathbf{k},t)$. Due to the relabeling of the sites, the momenta are transformed as
\begin{align}
\begin{split}
k_x &\rightarrow k_x
\\
k_y &\rightarrow k_y + k_x.
\end{split}
\end{align}
In particular, for $k_y=0$ we obtain the cut from $(0,0)$ to $(\pi,\pi)$ shown in the rightmost panel of Fig.~\ref{figSummary} (a). 

\begin{figure}[t!]
\centering
\epsfig{file=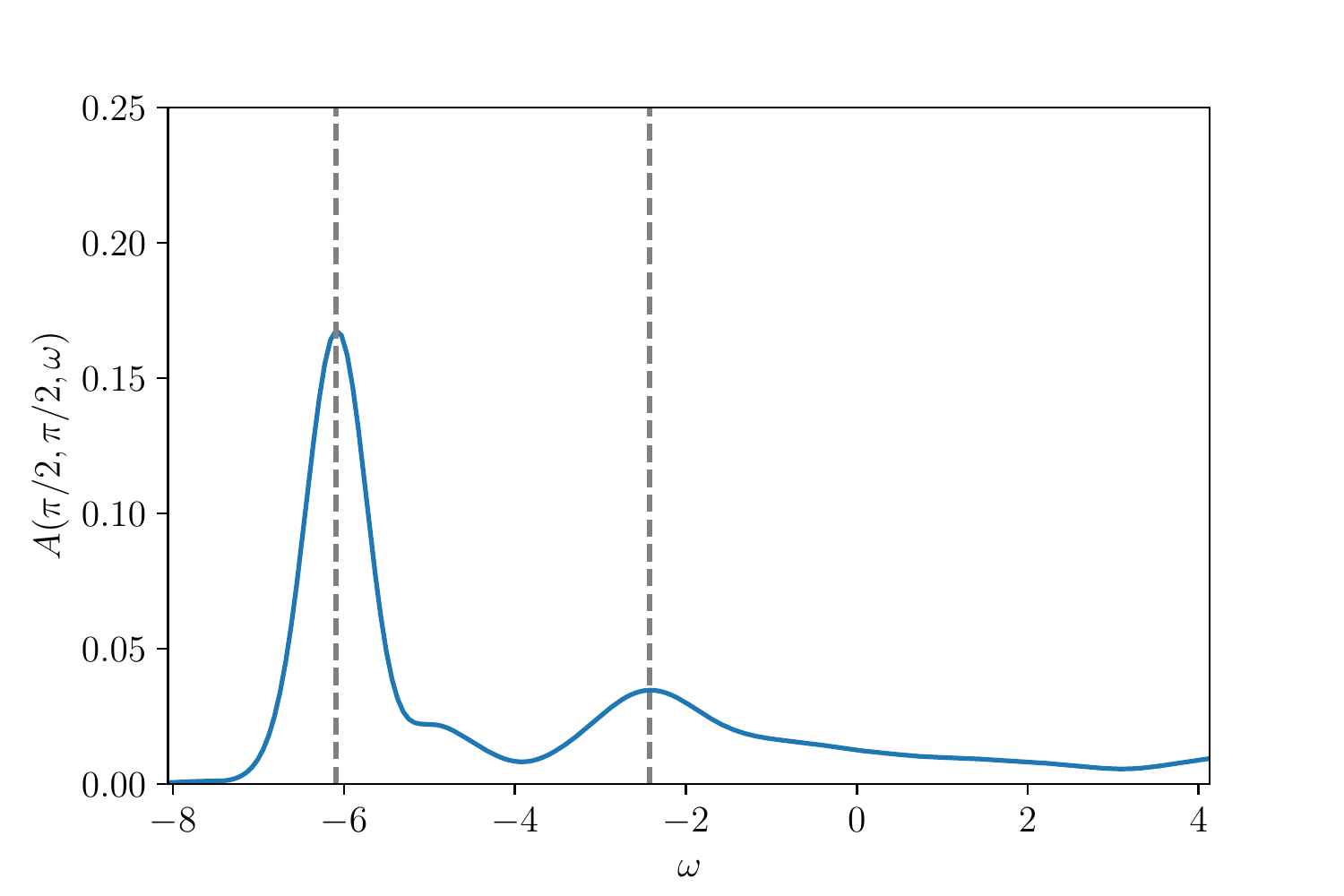, width=0.49\textwidth} 
\caption{\textsf{\textbf{First vibrational peak.}}  We show the frequency cut of the one-hole spectral function at the nodal point $\vec{k}=(\pi/2,\pi/2)$, for the same parameters as in Fig.~\ref{figSummary} (a) of the main text. The extracted positions of the ground state (first peak) and vibrationally excited (second peak) magnetic polaron are indicated by dashed lines.}
\label{figFirstVibrPeak}
\end{figure}

\emph{Extracting peak positions and quasiparticle weight.--}
From cuts at fixed momenta $\mathbf{k}$, the positions of the low energy peaks visible in Fig.~\ref{figSummary} (a) can be extracted. Fig.~\ref{figFirstVibrPeak} shows the corresponding cut at momentum $\mathbf{k}=(\pi/2,\pi/2)$ for $t/J=3$. 

\begin{figure}[t!]
\centering
\epsfig{file=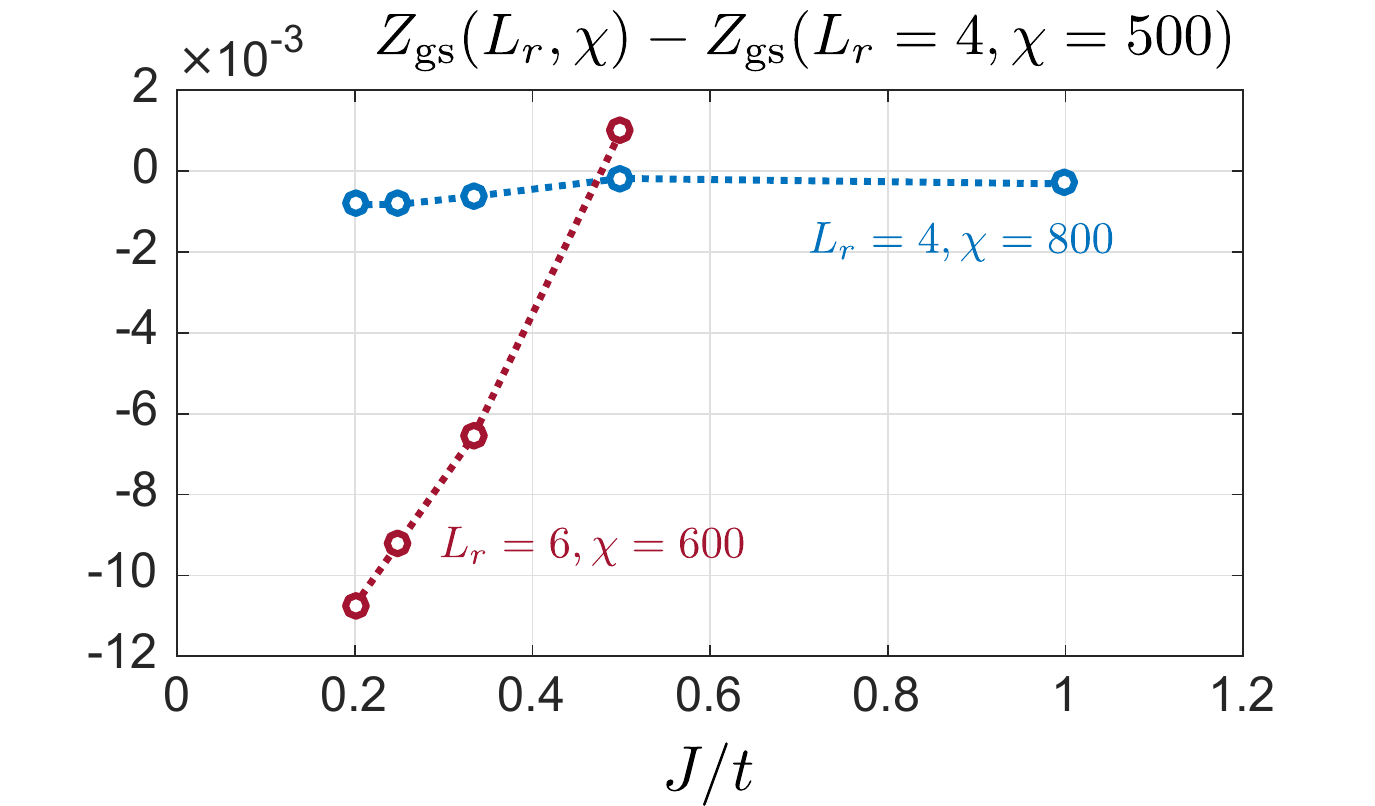, width=0.48\textwidth} $\quad$
\caption{\textsf{\textbf{Dependence of DMRG data on parameters.}}  We check how our DMRG calculations of the ground state quasiparticle weight $Z$ change when the bond dimension $\chi$ and the circumference $L_r$ of the cylinder are changed.}
\label{figZcDMRGconv}
\end{figure}

It can be shown analytically that the ground state quasiparticle weight $Z(\pi/2,\pi/2)$ shown in Fig.~\ref{figDeptJ} in the main text corresponds to the integral over the first peak in the spectral function. However, $Z(\pi/2,\pi/2)$ can be expressed more conveniently as the overlap:
\begin{equation}
Z(\pi/2,\pi/2) = \sum_\sigma \left| \bra{\psi_0^{\rm 1h}} \hat{c}_{(\pi/2,\pi/2),\sigma} \ket{\psi_0^{\rm 0h}} \right|^2.
\end{equation} 
We can thus obtain the quasiparticle weight directly from the ground states of the $t-J$ model without a hole, $\ket{\psi_0^{\rm 0h}}$, and with a single hole, $\ket{\psi_0^{\rm 1h}}$, without the need to calculate any time evolution. The ground state of a single hole in the $t-J$ model has momentum $\mathbf{k} = (\pi/2,\pi/2)$. Therefore we can further simplify the calculation by writing:
\begin{equation}
Z(\pi/2,\pi/2) = \sum_\sigma \sum_{\vec{j}} \left| \bra{\psi_0^{\rm 1h}} \hat{c}_{\vec{j},\sigma} \ket{\psi_0^{\rm 0h}} \right|^2.
\end{equation} 
Hence we only need to calculate the overlap of the one-hole ground state with a locally created hole at different positions $\vec{j}$: $\hat{c}_{\vec{j},\sigma} \ket{\psi_0^{\rm 0h}}$. In Fig.~\ref{figZcDMRGconv} we show how this procedure changes with the bond dimension $\chi$ and circumferences $L_r=4$ and $6$, for different $t/J$.

To extract $Z(\vec{k})$ for $\vec{k}$ different from the nodal points, we determine the height of the first peak and its full width at half maximum. We assume a Gaussian form and thus calculate $Z(\vec{k})$ as the corresponding integral over the Gaussian function.

\begin{figure}[t!]
\centering
\epsfig{file=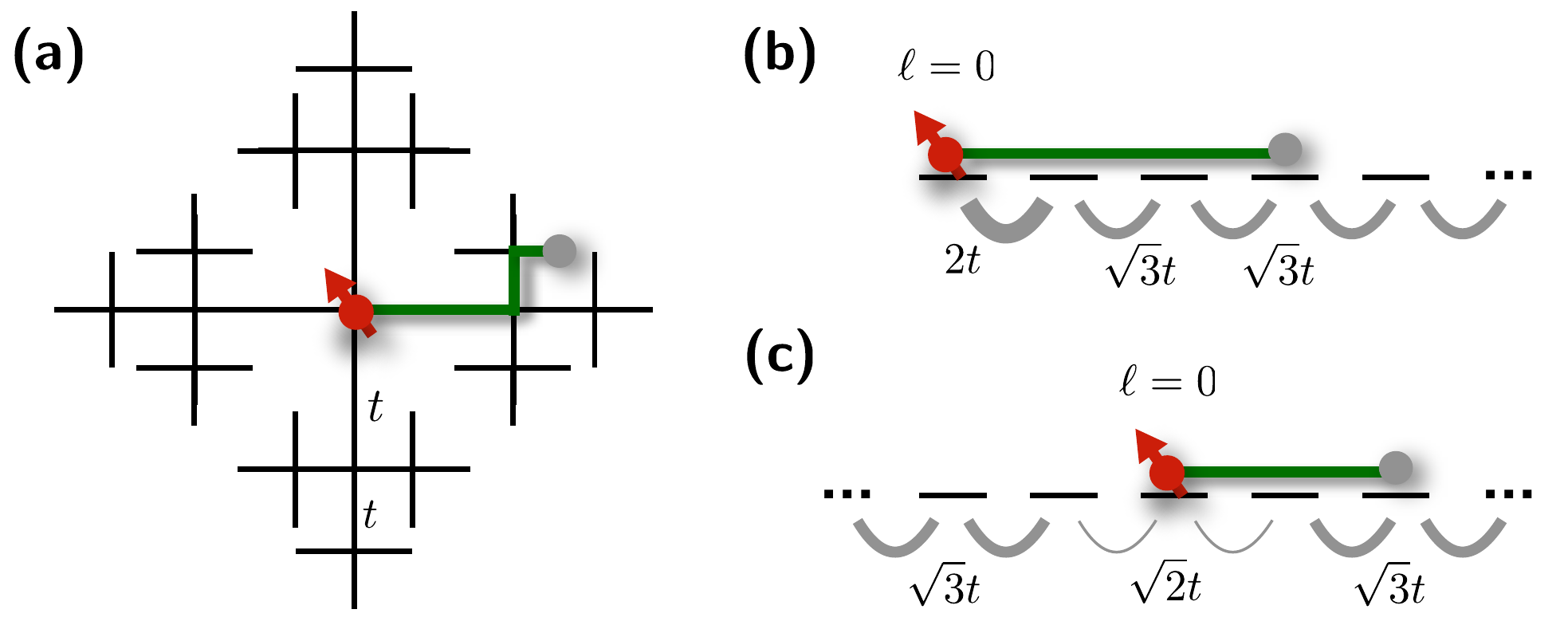, width=0.5\textwidth}
\caption{\textsf{\textbf{Spinon-chargon repulsion.}} (a) We consider a model where the charge fluctuations are described by string states, corresponding to the sites of a Bethe lattice. (b) Assuming a central-symmetric potential on the Bethe lattice, the one-particle hopping problem on the Bethe lattice, describing the fluctuating string, can be mapped to a semi-infinite one-dimensional chain with the indicated hopping amplitudes and a linear potential along the chain. (c) The problem in (b) can be related to an infinite one-dimensional problem with a mirror symmetry around the origin and \emph{reduced} hopping matrix elements $\sqrt{2} t$ to the central site, as compared to $\sqrt{3} t$ elsewhere. As described in the text, this inhomogeneous tunneling gives rise to a strong (of order $t$) microscopic spinon-chargon repulsion.}
\label{figSCrep}
\end{figure}

\textsf{\textbf{B) FSA ansatz.}}
We calculate the chargon, or string, contribution to the quasiparticle weight $Z_{\rm c}(J/t)$ using the FSA ansatz from Eq.~\eqref{eqFSA} in the main text, assuming that geometric string states are mutually orthogonal, $\bra{ \Sigma } \Sigma' \rangle \approx \delta_{\Sigma, \Sigma'}$. In addition, we simplify the effective string Hamiltonian, see Fig.~\ref{figSCrep} (a),
\begin{equation}
\H_\Sigma = -t \sum_{\langle \Sigma' , \Sigma \rangle} \bigl( \ket{\Sigma'} \bra{\Sigma} + \hc  \bigr) + \sum_\Sigma V_{\rm pot}(\Sigma) \ket{\Sigma}\bra{\Sigma}
\label{eqHSigmaLST}
\end{equation}
by making the linear string approximation \cite{Grusdt2019PRB}:
\begin{equation}
V_{\rm pot}(\Sigma) \approx V_{\rm LST}(\ell_\Sigma) = \frac{dE}{d\ell} \ell_\Sigma + g_0 \delta_{\ell_\Sigma,0} + \mu_{\rm h}.
\label{eqVLSTdef}
\end{equation}
Here $\ell_\Sigma$ denotes the length of string $\Sigma$ and the linear string tension is given by $dE/d\ell = 2 J \l C_2 - C_1 \r$, where $C_2$ ($C_1$) is the diagonal next-nearest neighbor (nearest neighbor) spin correlator $\langle \hat{\vec{S}}_{\vec{i}} \cdot \hat{\vec{S}}_{\vec{j}} \rangle$ in the undoped Heisenberg AFM. The additional point-like spinon-chargon attraction is given by $g_0 = - J \l C_3 - C_1 \r$ with $C_3$ the next-next nearest neighbor correlator. The zero-point energy $\mu_{\rm h} = J \l 1 + C_3 - 5 C_1 \r$ provides an overall energy offset; see Ref.~\cite{Grusdt2019PRB} for more details.  

Eq.~\eqref{eqHSigmaLST} describes a hopping problem on a Bethe lattice in the presence of a central-symmetric potential. Making use of all discrete rotational symmetries at the branches of the Bethe lattice, the problem can be reduced to a single particle in an effective semi-infinite one-dimensional lattice \cite{Bulaevskii1968,Grusdt2018PRX}, see Fig.~\ref{figSCrep} (b):
\begin{multline}
\H_{\rm eff} = \biggl[ - 2 t \ket{1} \bra{0}  - \sqrt{3} t \sum_{\ell = 1}^\infty \ket{\ell+1}\bra{\ell} + \hc \biggr] \\
+ \sum_{\ell=0}^\infty V_{\rm LST}(\ell) ~ \ket{\ell} \bra{\ell},
\label{eqHeffSemiInf}
\end{multline}
which can be solved exactly numerically. From the solution $\ket{\psi_\Sigma} = \sum_{\ell=0}^\infty \psi_\ell \ket{\ell}$, we obtain the chargon contribution to the quasiparticle weight as $Z_{\rm c} =|\psi_{\ell=0}|^2$. This result is plotted in Fig.~\ref{figDeptJ} of the main text. By diagonalization of Eq.~\eqref{eqHeffSemiInf} the excitation energy of the first vibrational state is also obtained, which we plot in Fig.~\ref{figGapVibrExct} as a function of $t/J$.
 
Now we discuss how $Z_{\rm c}(J/t)$ depends on $J/t$, focusing in particular on the asymptotic behavior when $J/t \to 0$. A naive mapping of Eq.~\eqref{eqHeffSemiInf} to the continuum limit $\ell \in \mathbb{R}_{>0}$ yields the Schr\"odinger equation \cite{Bulaevskii1968}
\begin{equation}
\left[ - \sqrt{3} t ~ \partial_\ell^2 + \frac{dE}{d\ell} \ell - E \right] \psi(\ell) = 0.
\label{eqSchrdgrCont}
\end{equation}
It is well known that the competition of the kinetic energy $\propto t$ and the linear string tension $dE/d\ell \propto J$ leads to an emergent average string length $L_\Sigma \propto (t/J)^{1/3}$ \cite{Bulaevskii1968}. Hence one would naively expect $Z_{\rm c} = |\psi_{\ell=0}|^2 \simeq L^{-1}_\Sigma = (J/t)^{1/3}$. When $J \simeq t$ are comparable in magnitude, such behavior is indeed observed \cite{Leung1995}. For most values shown in Fig.~\ref{figDeptJ} the residue of the magnetic polaron $Z(J/t) \simeq (J/t)^\alpha$ can be approximated by a power-law, and for $t \simeq J$ the exponent $\alpha < 1$ is significantly below one. This is consistent with the naive expectations above.

\begin{figure}[t!]
\centering
\epsfig{file=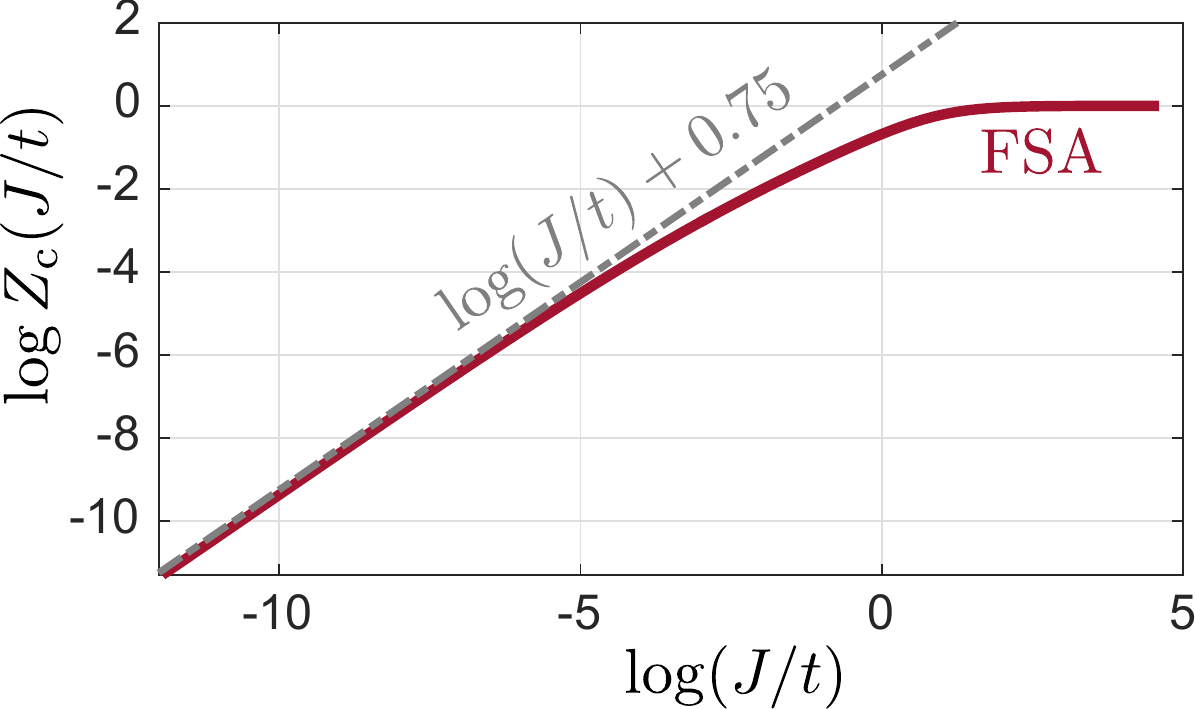, width=0.38\textwidth} $\qquad$
\caption{\textsf{\textbf{Chargon contribution the quasiparticle weight.}} $Z_{\rm c}(J/t)$ is calculated from the model in Eq.~\eqref{eqHeffSemiInf}, based on the frozen spin approximation (FSA), and shown in a double-logarithmic plot. In the strong coupling regime $J \ll t$ the asymptotic power-law $Z_{\rm c}(J/t) \propto J/t$ is obtained as a result of spinon-chargon repulsion.}
\label{figZcFSA}
\end{figure}

In Fig.~\ref{figZcFSA} we calculate $Z_{\rm c}(J/t)$ from the model in Eq.~\eqref{eqHeffSemiInf} and show the result in a log-log plot. For $J \ll t$, in the strong coupling regime, our result demonstrates that $Z_{\rm c}(J/t) \simeq J/t$, i.e. asymptotically the exponent of the power law approaches $\alpha=1$. This indicates that additional spinon-chargon repulsion must be present in the effective continuum model \eqref{eqSchrdgrCont}, which enhances the exponent from $\alpha=1/3$ to $\alpha=1$ at strong couplings. 

The origin of additional spinon-chargon repulsion can be understood by mapping the semi-infinite one-dimensional problem \eqref{eqHeffSemiInf} to the even-parity sector of the following infinite problem, see Fig.~\ref{figSCrep} (c),
\begin{multline}
\H_{\rm eff}' = - \sqrt{3} t  \sum_{\ell=- \infty}^\infty \biggl[  (1 - \delta_{\ell,0} - \delta_{\ell,-1}) ~ \ket{\ell+1} \bra{\ell} + \hc \biggr] \\
- \sqrt{2} t  \Bigl[  \ket{1}\bra{0} + \ket{0}\bra{-1} + \hc \Bigr] + \sum_{\ell=- \infty}^\infty V_{\rm LST}(\ell) ~ \ket{\ell} \bra{\ell}.
\label{eqHeffInf}
\end{multline}
One can confirm that every even parity eigenstate $\phi_{-\ell} = \phi_\ell$ of $\H_{\rm eff}'$ corresponds to an eigenstate of $\H_{\rm eff}$ given by $\psi_\ell = \sqrt{2} \phi_\ell$ for $\ell > 0$ and $\psi_{\ell=0} = \phi_{\ell = 0}$, with the same eigenenergy. 
 
Eq.~\eqref{eqHeffInf} describes a single particle hopping in an infinite chain in the presence of a central-symmetric confining potential. Around the origin, the tunneling amplitudes are reduced from $\sqrt{3} t$ in the bulk to $\sqrt{2} t$. This reduces the zero-point kinetic energy from $- 2 \sqrt{3} t$ in the bulk to $- 2 \sqrt{2} t$ in the origin, corresponding to a localized repulsive potential with a strength of the order $2 (\sqrt{3} - \sqrt{2}) t = 0.64 t$. At strong couplings this repulsion overcomes the spinon-chargon attraction $\propto g_0 \propto J$ included in $V_{\rm LST}(\ell)$, see Eq.~\eqref{eqVLSTdef}, and leads to the formation of a node at $\ell = 0$ in the string wavefunction when $J \ll t$. This effect is not included in the naive continuum theory Eq.~\eqref{eqSchrdgrCont}, and explains why we observe $Z_{\rm c}(J/t) \propto J/t$ asymptotically when $J/t \to 0$.
 
The effective local spinon-chargon repulsion $\propto t$ is determined by the connectivity of the Bethe lattice defining the string basis. For chargons moving only along one dimension, as realized for example in the mixed-dimensional $t-J$ model \cite{Grusdt2018SciPost}, this additional repulsion is absent. In such settings, a different power-law is reached asymptotically at strong couplings.\\

\textsf{\textbf{C) Spinon-chargon trial wavefunction.}}
We use the trial wavefunction from Eq.~\eqref{eqDefMesonWvfct} to calculate the quasiparticle weight $Z(\vec{k})$, see Eq.~\eqref{eqDefZsc}. To evaluate the expression in Eq.~\eqref{eqDefZsc} we use Metropolis Monte Carlo. 

\emph{Sampling.--} 
We sample Fock configurations $\ket{\alpha}$ of the fermionic spinons $\f_{\vec{j},\sigma}$, and two sets of string configurations $\Sigma$ and $\Sigma'$ from the following positive-definite distribution:
\begin{multline}
\rho_J(\alpha,\Sigma,\Sigma') = |\psi_\Sigma|  |\psi_{\Sigma'}|  |\bra{\Psi_{\rm MF}(\vec{k},\sigma)}  \alpha_{\Sigma} \rangle | \\
\times   |\bra{\Psi_{\rm MF}(\vec{k},\sigma)}  \alpha_{\Sigma'} \rangle |. 
\end{multline}
Here $\ket{\Psi_{\rm MF}(\vec{k},\sigma)} = \f_{\vec{k},\sigma,-} \ket{\Psi_{\rm MF}^{\rm SF+N}}$ is the mean-field state with one extra spinon excitation of momentum $\vec{k}$ in the lower spinon band ($\mu=-$). We further introduced the squeezed space Fock configurations
\begin{equation}
\ket{\alpha_\Sigma} = \hat{G}_\Sigma^\dagger \ket{\alpha},
\end{equation}
where the operator $\hat{G}_\Sigma$ also appears in the definition of the trial wavefunction \eqref{eqDefMesonWvfct} and creates a state with a geometric string $\Sigma$ starting at the position of the hole in the Fock state to its right; see Eq.~\eqref{eqDefG}.

To calculate $Z(\vec{k})$ we introduce a completeness relation of one-hole Fock states, $\sum_\alpha \ket{\alpha}\bra{\alpha} = \hat{1}_{\rm 1h}$, and use momentum conservation. This leads to
\begin{multline}
Z(\vec{k}) = \frac{1}{\sqrt{\mathcal{N}_0 \mathcal{N}_1}} \sum_\sigma \sum_{\vec{j}} \frac{e^{i \vec{k} \cdot \vec{j}}}{L} \sum_\Sigma \psi_\Sigma^* \sum_\alpha \\
\times \bra{\Psi_{\rm MF}(\vec{k},\sigma)}  \hat{\mathcal{P}}_{\rm GW}  \ket{\alpha_\Sigma} \bra{\alpha} \f_{\vec{j},\sigma}   \hat{\mathcal{P}}_{\rm GW}  \ket{\Psi_{\rm MF}^{\rm SF+N}},
\label{eqZkDeriv1}
\end{multline}
with the normalizations
\begin{flalign}
\mathcal{N}_0 &= \bra{\Psi_{\rm sc}(\vec{k})} \Psi_{\rm sc}(\vec{k}) \rangle, \\
\mathcal{N}_1 &= \bra{\Psi_{\rm MF}^{\rm SF+N}(\vec{k})}  \hat{\mathcal{P}}_{\rm GW} \ket{ \Psi_{\rm MF}^{\rm SF+N}(\vec{k}) }.
\end{flalign}
Note that for the trial wavefunction it holds $\ket{\Psi_{\rm sc}(\vec{k})} \equiv \ket{\Psi_{\rm sc}(\vec{k}+\vec{K})}$ for reciprocal lattice vectors $\vec{K}$.

The first Gutzwiller projection in Eq.~\eqref{eqZkDeriv1} can be dropped, since it acts on a Fock configuration to the right. The second Gutzwiller projection can be handled in a similar way, by writing
\begin{equation}
 \frac{1}{L} \sum_{\vec{j}} e^{- i \vec{k} \cdot \vec{j}}  \hat{\mathcal{P}}_{\rm GW} \fd_{\vec{j},\sigma} \ket{\alpha}  = \frac{L^2/2}{L} e^{- i \vec{k} \cdot \vec{j}_\alpha^{\rm h}} \ket{\tilde{\alpha}}
\end{equation}
where $L^2/2 \ket{\tilde{\alpha}} =  \hat{\mathcal{P}}_{\rm GW} \fd_{\vec{j},\sigma} \ket{\alpha}$ and $\vec{j}_\alpha^{\rm h}$ denotes the position of the hole in the Fock configuration $\ket{\alpha}$. The state $\ket{\tilde{\alpha}}$ is thus obtained from $\ket{\alpha}$ by adding a fermion at site $\vec{j}_\alpha^{\rm h}$. The extra factor $L^2/2 = N_\sigma$ is equal to the number of spins $\sigma$ in Fock state $\ket{\alpha}$, and arises when relating the properly normalized first ($\ket{\tilde{\alpha}}$) and second ($ \hat{\mathcal{P}}_{\rm GW} \fd_{\vec{j},\sigma} \ket{\alpha}$) quantized many-body states.

Combining the results above, we can write $Z(\vec{k})$ as
\begin{widetext}
\begin{equation}
Z(\vec{k}) = \sum_\sigma \sum_{\Sigma,\Sigma'} \sum_\alpha \rho_J(\alpha,\Sigma,\Sigma') \frac{\delta_{\Sigma,\Sigma'} ~ \psi_\Sigma^*  ~ \bra{\Psi_{\rm MF}(\vec{k},\sigma)} \alpha_\Sigma \rangle  ~ e^{i \vec{k} \cdot \vec{j}_\alpha^{\rm h}} ~ \bra{\tilde{\alpha}} \Psi_{\rm MF}^{\rm SF+N}  \rangle }{\rho_J(\alpha,\Sigma,\Sigma') ~ \sqrt{2}}  ~ \mathcal{N}^{-1/2},
\label{eqZkForSampling}
\end{equation}
where
\begin{equation}
\mathcal{N} =  \biggl[ \sum_{\Sigma,\Sigma'} \sum_\alpha \rho_J(\alpha,\Sigma,\Sigma') \frac{\psi^*_\Sigma~  \psi_{\Sigma'} ~ \bra{ \Psi_{\rm MF}(\vec{k},\sigma) } \alpha_\Sigma \rangle ~ \langle \alpha_{\Sigma'} \ket{\Psi_{\rm MF}(\vec{k},\sigma)}}{\rho_J(\alpha,\Sigma,\Sigma') }  \biggr]
 \biggl[ \sum_{\Sigma,\Sigma'} \sum_\alpha \rho_J(\alpha,\Sigma,\Sigma') \frac{\delta_{\Sigma,\Sigma'} ~ | \psi_\Sigma|^2 ~  | \bra{ \Psi_{\rm MF}^{\rm SF+N} } \tilde{\alpha}_\Sigma \rangle|^2 }{\rho_J(\alpha,\Sigma,\Sigma') }  \biggr].
\end{equation}
\end{widetext}
This is the expression we used to perform a Metropolis Monte Carlo procedure, sampling from the distribution $\rho_J(\alpha,\Sigma,\Sigma')$. Note that the required Fock-state overlaps can be straightforwardly evaluated, see e.g. Refs.~\cite{Gros1989,Piazza2015}.

\begin{figure}[t!]
\centering
\epsfig{file=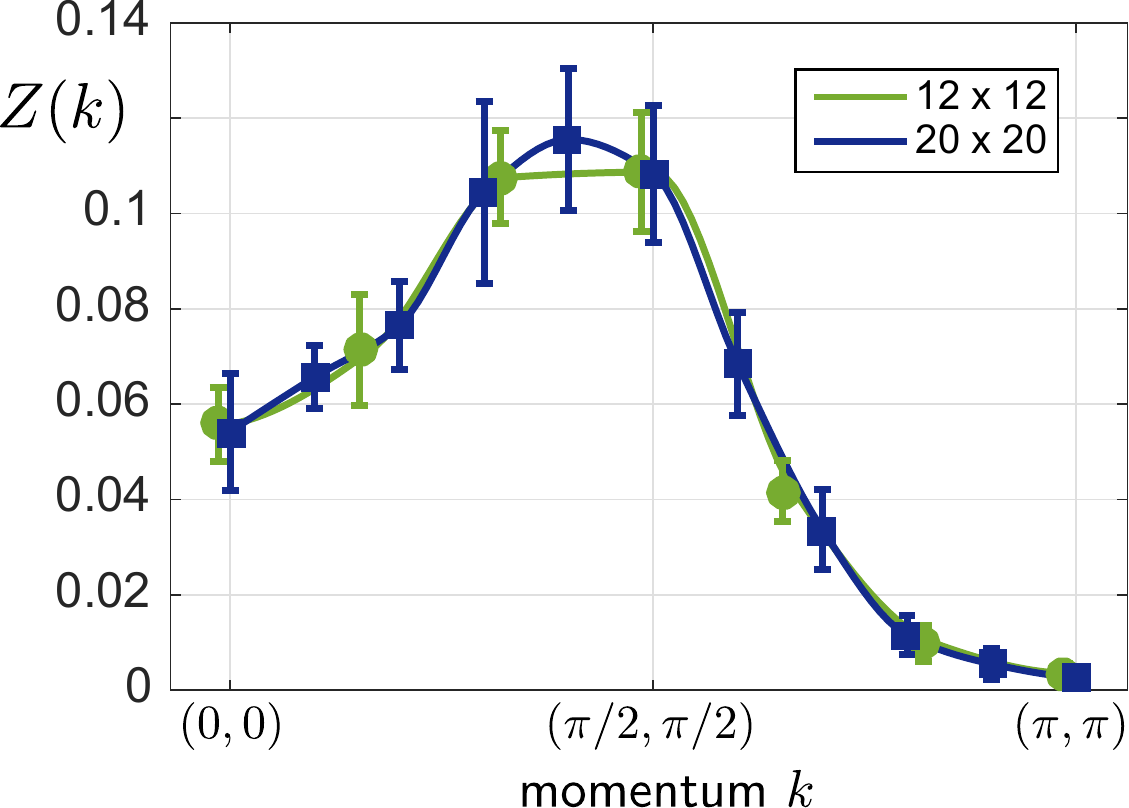, width=0.42\textwidth} $\quad$
\caption{\textsf{\textbf{Size-dependence of the quasiparticle weight.}} $Z(k)$ is shown along the diagonal cut from $(0,0)$ to $(\pi,\pi)$. We evaluated the spinon-chargon trial wavefunction in a $12 \times 12$ and $20 \times 20$ system. No significant finite-size dependence can be observed. We set $t/J=3$, $B_{\rm st} / J_{\rm eff} = 0.44$ and $\Phi = 0.4 \pi$. Solid lines are guides to the eye only; data points for $12 \times 12$ are slightly offset horizontally for better visibility.}
\label{figZfiniteSize}
\end{figure}

\emph{Parameter dependence.--} 
We worked with Eq.~\eqref{eqZkForSampling} to check how the quasiparticle residue of the trial wavefunction depends on various parameters. In Fig.~\ref{figZfiniteSize} we compare $Z(\vec{k})$ along a diagonal cut through the magnetic Brillouin zone for different system sizes. We show that the result does not change significantly when increasing the system size from $12 \times 12$ to $20 \times 20$, indicating that finite-size scaling of our results obtained in $12 \times 12$ systems is not necessary. 

\begin{figure}[t!]
\centering
\epsfig{file=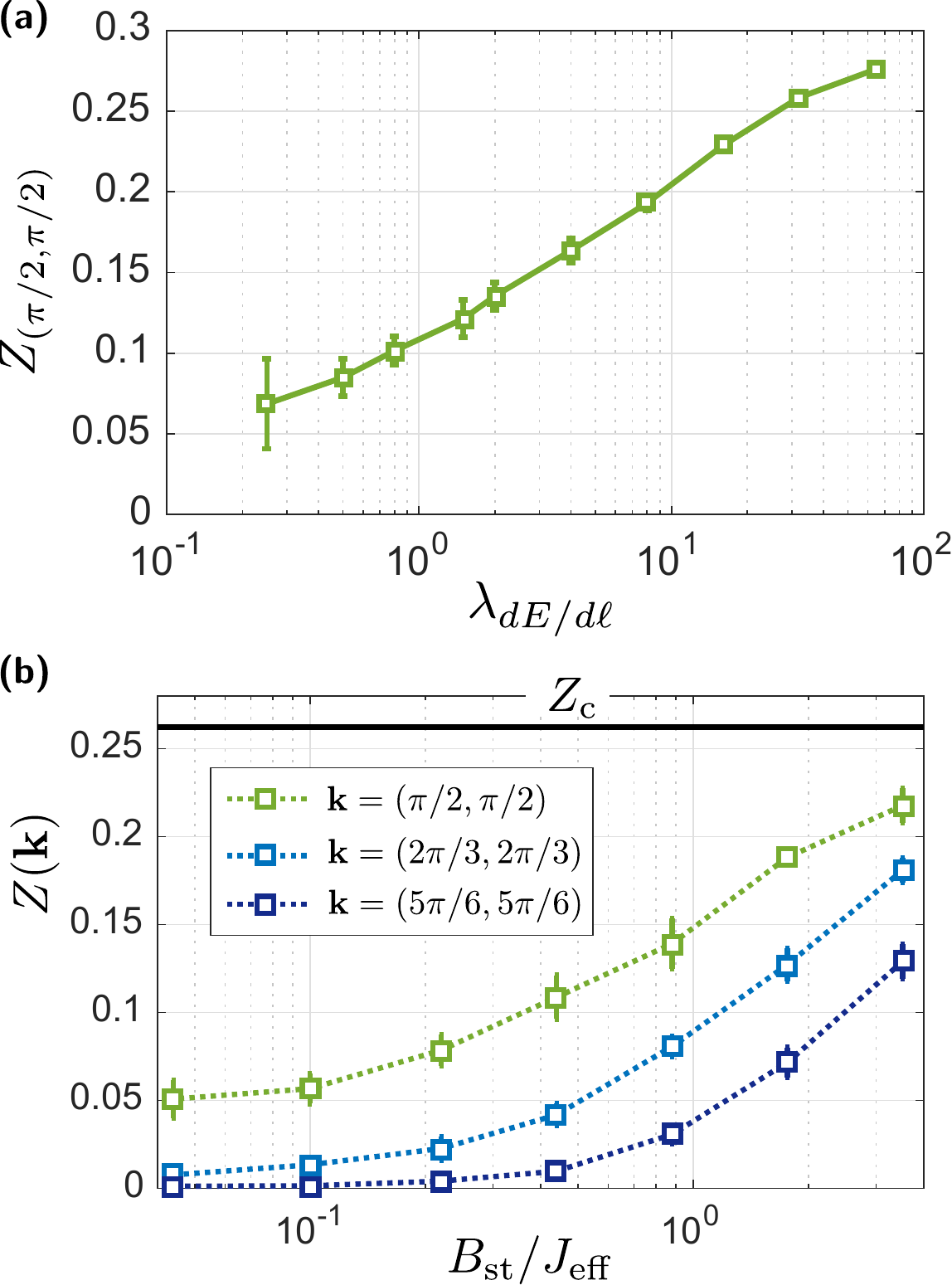, width=0.42\textwidth} $\quad$
\caption{\textsf{\textbf{Dependence of the quasiparticle weight on string tension and staggered magnetic field.}} (a) We calculate the quasiparticle residue $Z(\pi/2,\pi/2)$ from the trial wavefunction for different string length distributions. The latter are determined by the linear string tension $dE / d\ell = \lambda_{dE/d\ell} 2 J (C_{\vec{e}_x+\vec{e}_y} - C_{\vec{e}_x})$ used in the FSA ansatz; Large $\lambda_{dE/d\ell} \gg 1$ corresponds to short strings, and vice-versa. We use parameters $B_{\rm st} / J_{\rm eff}=0.44$ and $\Phi=0.4 \pi$ in a $12 \times 12$ system. For $t/J=3$ the variational energy is minimized for $\lambda_{dE/d\ell}$ between $10^0$ to $10^1$, see Ref.~\cite{Grusdt2019PRB}. (b) The quasiparticle weight is calculated for different values of the staggered magnetic field $B_{\rm st}/J_{\rm eff}$ in the trial wavefunction Eq.~\eqref{eqDefMesonWvfct}. For large staggered fields, the chargon, or string, contribution $Z_{\rm c}$ is approached. For weak staggered fields a strong momentum-dependence is observed. We set $t=3J$, $\Phi=0.4 \pi$ and worked in a $12 \times 12$ system.}
\label{figDepdEdl}
\end{figure}

In Fig.~\ref{figDepdEdl} (a) we show how $Z(\vec{k})$ depends on the linear string tension, which determines how tightly the chargon is bound to the spinon. We rescaled the FSA expression for $dE/d\ell$, introduced below Eq.~\eqref{eqVLSTdef}, by a factor $\lambda_{dE/d\ell}$. The data shown in the main text corresponds to $\lambda_{dE/d\ell}=1$. Assuming tighter spinon-chargon confinement leads to significantly increased spectral weight. This may explain why the result by the trial wavefunction in Fig.~\ref{figDeptJ} of the main text showed too small quasiparticle residues as compared to the numerically obtained results starting from first principles. Note that similar indications for tighter spinon-chargon confinement have been obtained in studies of the variational energy \cite{Grusdt2019PRB}.

In Fig.~\ref{figDepdEdl} (b) we show how $Z(\vec{k})$ depends on the staggered field $B_{\rm st}/J_{\rm eff}$ characterizing the undoped trial wavefunction. In the limit $B_{\rm st}/J_{\rm eff} \to 0$ different behavior is observed for momenta $\vec{k}$ within and outside the magnetic Brillouin zone. For larger values of $B_{\rm st}/J_{\rm eff}$, larger quasiparticle residues are obtained. This may also play a role for explaining the deviations observed in Fig.~\ref{figDeptJ} of the main text between the trial wavefunction and numerical approaches. \\

\textsf{\textbf{D) Mean-field approximation and spinon statistics.}}
In this section, we describe the phenomenology of the quasiparticle weight $Z_{\rm s}(\vec{k})$ expected from two different mean-field descriptions. Only the fermionic theory is consistent with numerical results.

\emph{Fermionic spinons.--}
In this approach, we assume that the constituting spinons $\f_{\vec{j},\sigma}$ introduced in Eq.~\eqref{eqDefSpnonHolon} obey fermionic statistics. In our derivation of the mean-field expression $Z^{\rm MF}_{\rm s}(\vec{k})$ in Eq.~\eqref{eqZMF} of the main text, we first assume that only $\Sigma = 0$ leads to a non-vanishing contribution:
\begin{multline}
\bra{ \Psi_{\rm sc}(\vec{k}) } \hat{f}_{\vec{k},\sigma} \hat{\mathcal{P}}_{\rm GW} \ket{\Psi_{\rm MF}^{\rm SF+N}} \approx \psi_{\Sigma=0}^* \sum_{\vec{j}^{\rm s}}
\frac{u_{\vec{k},\sigma,-}^{(\vec{j}^s)} e^{-i \vec{k} \cdot \vec{j}^s}}{L / \sqrt{2}} \\
\times \bra{\Psi_{\rm MF}^{\rm SF+N}} \fd_{\vec{j}^s,\sigma} ~ \hat{\mathcal{P}}_{\rm GW} ~ \hat{f}_{\vec{k},\sigma} \hat{\mathcal{P}}_{\rm GW} \ket{\Psi_{\rm MF}^{\rm SF+N}} \\
= \psi_{\Sigma=0}^*  ~  \bra{\Psi_{\rm MF}^{\rm SF+N}}   \fd_{\vec{k},\sigma,-}  ~ \hat{\mathcal{P}}_{\rm GW} ~ \hat{f}_{\vec{k},\sigma} \hat{\mathcal{P}}_{\rm GW} \ket{\Psi_{\rm MF}^{\rm SF+N}}.
\end{multline}
In the second step, we used 
\begin{equation}
\fd_{\vec{k},\sigma,\mu} = \sum_{\vec{j}} \frac{e^{ - i \vec{k} \cdot \vec{j}}}{L/\sqrt{2}}  ~  u^{(\vec{j})}_{\vec{k},\sigma,\mu}  ~  \fd_{\vec{j,\sigma}}.
\end{equation}
Next we drop the Gutzwiller projectors and approximate
\begin{multline}
 \bra{\Psi_{\rm MF}^{\rm SF+N}}   \fd_{\vec{k},\sigma,-}  ~ \hat{\mathcal{P}}_{\rm GW} ~ \hat{f}_{\vec{k},\sigma} \hat{\mathcal{P}}_{\rm GW} \ket{\Psi_{\rm MF}^{\rm SF+N}} \\
 \approx  \bra{\Psi_{\rm MF}^{\rm SF+N}}   \fd_{\vec{k},\sigma,-}   ~ \hat{f}_{\vec{k},\sigma} \ket{\Psi_{\rm MF}^{\rm SF+N}},
\end{multline}
which yields the mean-field result Eq.~\eqref{eqZMF} in the main text. Note that $\f_{\vec{k}+\vec{K},\sigma,-} \equiv \f_{\vec{k},\sigma,-}$, where $\vec{K}$ is the reciprocal lattice vector, but $\f_{\vec{k}+ \vec{K},\sigma} \neq \f_{\vec{k},\sigma}$.

We use the following identities
\begin{flalign}
\fd_{\vec{k},\sigma} &= \lambda_{\vec{k}}^+ \fd_{\vec{k},\sigma,+} + \lambda_{\vec{k}}^- \fd_{\vec{k},\sigma,-}, \\
\fd_{\vec{k}+\vec{K},\sigma} &= \lambda_{\vec{k}+\vec{K}}^+ \fd_{\vec{k},\sigma,+} + \lambda_{\vec{k}+\vec{K}}^- \fd_{\vec{k},\sigma,-}, \\
\end{flalign}
where $\vec{k} \in {\rm MBZ}$ and the factors $\lambda^{\pm}$ are given by
\begin{flalign}
\lambda_{\vec{k}}^\mu &= \frac{1}{\sqrt{2}} \l u^{(A)}_{\vec{k},\sigma,\mu} + u^{(B)}_{\vec{k},\sigma,\mu} \r^* \\
\lambda_{\vec{k}+\vec{K}}^\mu &= \frac{1}{\sqrt{2}} \l u^{(A)}_{\vec{k},\sigma,\mu} - u^{(B)}_{\vec{k},\sigma,\mu} \r^*;
\end{flalign}
$A$ and $B$ denote sites $\vec{j}$ from the $A$ and $B$ sublattice, respectively, and $\mu=\pm$ is the band index. This leads to the result in Eq.~\eqref{eqZsMF} in the main text.

\emph{Bosonic spinons.--}
In this approach, we assume that the constituting spinons $\f_{\vec{j},\sigma}$ introduced in Eq.~\eqref{eqDefSpnonHolon} obey bosonic statistics. To avoid confusion, we replace $\f_{\vec{j},\sigma}$ by Schwinger bosons $\a_{\vec{j},\sigma}$ in the following. First we introduce the bosonic analogue of the trial wavefunction Eq.~\eqref{eqDefMesonWvfct} in the main text. We start from the usual Holstein-Primakoff expansion, $\hat{S}^z_{\vec{j}} = (-1)^{\vec{j}} ( \nicefrac{1}{2} - \ad_{\vec{j}} \a_{\vec{j}})$, where $(-1)^{\vec{j}} = (-1)^{j_x+j_y}$. This corresponds to a mean-field expansion around the condensates $\a_{\vec{j}_A,\uparrow}, \a_{\vec{j}_B,\downarrow} \to \alpha_{\vec{j}_A,\uparrow}, \alpha_{\vec{j}_B,\downarrow}=1$. Here $\vec{j}_{A,B}$ denote lattice sites from the $A$ and $B$ sublattices respectively and $\alpha_{\vec{j},\sigma}$ denotes coherent state amplitudes. The operators $\a_{\vec{j}_A} = \a_{\vec{j}_A,\downarrow}$ and $\a_{\vec{j}_B} = \a_{\vec{j}_B,\uparrow}$ describe spin-flips of the anti-ferromagnet, giving rise to collective spin-wave excitations which carry spin $S=1$. 

The fluctuations $\b_{\vec{j},\uparrow} = \a_{\vec{j}_A,\uparrow}$ for $\vec{j}=\vec{j}_A \equiv \vec{j}_\uparrow$ and $\b_{\vec{j},\downarrow} = \a_{\vec{j}_B,\downarrow}$ for $\vec{j}=\vec{j}_B \equiv \vec{j}_\downarrow$ around the condensate describe vacancies in the classical N\'eel state around which we expand in the Holstein-Primakoff approach. Hence they correspond to spin-$1/2$ excitations, and we interpret them as bosonic spinons. This leads us to the following form of spinon-chargon trial states,
\begin{equation}
\ket{\Psi_{\rm sc}(\vec{k})} = \sum_{\vec{j}^s_\sigma}  \frac{e^{i \vec{k} \cdot \vec{j}^s}}{L / \sqrt{2}} \\
  \sum_\Sigma \psi_\Sigma ~ \hat{G}_\Sigma ~ \hat{\mathcal{P}}_{\rm GW} ~ \b_{\vec{j}^s,\sigma} \ket{\Psi_{\rm MF}^{\rm SB}},
\end{equation}
c.f. Eq.~\eqref{eqDefMesonWvfct}. Here $\sum_{\vec{j}^s_\sigma}$ denotes a sum over all sites from sublattice $A$ for $\sigma=\uparrow$ and $B$ for $\sigma=\downarrow$. As in the fermionic case, it holds $\ket{\Psi_{\rm sc}(\vec{k}+\vec{K})} = \ket{\Psi_{\rm sc}(\vec{k})}$ up to an overall phase and for reciprocal lattice vectors $\vec{K}$. The bosonic mean-field description of the half-filled Heisenberg AFM is given by
\begin{equation}
\ket{\Psi_{\rm MF}^{\rm SB}} = \prod_{\vec{j}} \bigl( \ket{ \alpha_{\vec{j}_A,\uparrow} = 1 } \ket{ \alpha_{\vec{j}_B,\downarrow} = 1 } \bigr) \otimes \ket{\Psi_{\rm fluc}},
\label{eqDefPsiMFSB}
\end{equation}
where $\prod_{\vec{j}}$ is a product over all unit cells $\vec{j}$, each consisting of a site $\vec{j}_A$ from the $A$- and a site $\vec{j}_B$ from the $B$-sublattice. $\ket{\Psi_{\rm fluc}}$ denotes the bosonic Gaussian state of fluctuations in the Hilbertspace of $\a_{\vec{j}_A,\downarrow}$ and $\a_{\vec{j}_B,\uparrow}$ Schwinger bosons; $\ket{\alpha_{\vec{j},\sigma}}$ denotes a coherent state of Schwinger bosons with complex amplitude $\alpha_{\vec{j},\sigma}$. 

In the calculation of the quasiparticle residue within the mean-field theory we assume, as in the fermionic case, that only the trivial string configuration $\Sigma=0$ contributes. Dropping the Gutzwiller projection leads to the following bosonic mean-field expression,
\begin{equation}
Z^{\rm MF,B}(\vec{k}) = Z_{\rm c} \sum_\sigma | \bra{\Psi_{\rm MF}^{\rm SB}} \ad_{\vec{k},\eta_\sigma,\sigma} \a_{\vec{k},\sigma} \ket{\Psi_{\rm MF}^{\rm SB}} |^2,
\label{eqZMFB}
\end{equation}
c.f. Eq.~\eqref{eqZMF}. Here we introduced
\begin{equation}
\a_{\vec{k},\mu,\sigma} = \frac{\sqrt{2}}{L} \sum_{\vec{j}_\mu} e^{i \vec{k} \cdot \vec{j}_\mu} \a_{\vec{j}_\mu,\sigma},
\end{equation}
for $\mu=A,B$ and defined
\begin{equation}
\eta_\sigma=
\begin{cases}
A, \quad \sigma=\uparrow\\
B, \quad \sigma=\downarrow.
\end{cases} 
\end{equation}
Hence $\ad_{\vec{k},\eta_\sigma,\sigma}$ only involves the spinon operators $\b_{\vec{j},\sigma}$ but not the spin-flip operators $\a_{\vec{j}}$. 

To calculate $Z^{\rm MF,B}(\vec{k})$ in Eq.~\eqref{eqZMFB} we note that the Schwinger-boson mean-field state \eqref{eqDefPsiMFSB} can be written 
\begin{equation}
\ket{\Psi_{\rm MF}^{\rm SB}} =  \ket{ \alpha_{\vec{k}=0,A,\uparrow} }  \ket{ \alpha_{\vec{k}=0,B,\downarrow} }  \otimes \ket{\Psi_{\rm fluc}},
\end{equation}
with coherent amplitudes $\alpha_{\vec{k}=0,A,\uparrow} = \alpha_{\vec{k}=0,B,\downarrow} = \nicefrac{L}{\sqrt{2}}$; i.e. the free spinons condense at $\vec{k} = 0$ in the mean-field theory. This condensate is the bosonic counterpart of the Fermi sea formed by constituting spinons in the fermionic mean-field theory. 

The rest of the bosonic mean-field calculation is straightforward. Because $\a_{\vec{k},\mu,\sigma} = \a_{\vec{k}+\vec{K},\mu,\sigma}$ up to an overall phase for the reciprocal lattice vector $\vec{K} = (\pi,\pi)$, we find that 
\begin{equation}
Z^{\rm MF,B}(\vec{k}) = Z_{\rm c} \frac{L^2}{2} \biggl( \delta(\vec{k}) + \delta(\vec{k} - \vec{K}) \biggr).
\end{equation}
This directly leads to 
\begin{equation}
Z_{\rm s}^{\rm MF,B}(\vec{k}) \propto \bigl( \delta(\vec{k}) + \delta(\vec{k} - \vec{\pi}) \bigr), \quad \vec{\pi} = (\pi,\pi).
\label{eqZsMFB}
\end{equation}
Instead of the Fermi sea revealed in the fermionic mean-field theory, we expect two delta-distributions at $\vec{k}=(0,0)$ and $\vec{k}=(\pi,\pi)$. Beyond the mean-field ansatz, the Gutzwiller projection is expected to lead to substantial broadening of these delta-function peaks. While this may explain some of the numerical observations, it cannot explain the striking suppression of spectral weight around $\vec{k} = (\pi,\pi)$ observed numerically. While it is difficult to rule out bosonic descriptions of spinons completely, we conclude that strong interactions between the bosons would be required to explain the observed distribution of spectral weight across the Brillouin zone.

\newpage

\textsf{\textbf{\large References}}
\vspace{-1.5cm}

~ \\
\textsf{\textbf{Acknowledgements}}\\
\textbf{General:} The authors thank Z.X. Shen, S. Sachdev, A. Vishwanath, T. Giamarchi, U. Schollw\"ock, C. Hubig, I. Cirac, R. Verresen, D. Chowdhury, M. Punk, Y. Wang, A. Rosch, I. Bloch, M. Endres, G. Salomon, D. Greif, W. Bakr, P. Schauss for fruitful discussions. We additionally want to thank R. Verresen for sharing parts of the numerical code.\\
\textbf{Funding:} The authors acknowledge support from the Technical University of Munich - Institute for Advanced Study, funded by the German Excellence Initiative and the European Union FP7 under grant agreement 291763, the Deutsche Forschungsgemeinschaft (DFG, German Research Foundation) under Germany's Excellence Strategy EXC-2111- 390814868, DFG grant No. KN1254/1-1, DFG TRR80 (Project F8), the Studienstiftung des deutschen Volkes, Harvard-MIT CUA, AFOSR-MURI: Photonic Quantum Matter (award FA95501610323), DARPA DRINQS program (award D18AC00014). FP acknowledges support by the European Research Council (ERC) under the European Unions Horizon 2020 research and innovation program (grant agreement No. 771537), the DFG Research Unit FOR 1807 through grants no. PO 1370/2- 1, TRR80, and the Deutsche Forschungsgemeinschaft (DFG, German Research Foundation) under Germanys Excellence Strategy EXC-2111- 390814868.\\
%
%
\textbf{Competing interests:}
The authors declare no competing interests.\\
\textbf{Data availability:} 
The data that support the findings of this study are available from the corresponding author upon reasonable request.

\end{document}